\documentclass[
               twocolumn,
               noshowpacs,          
               nopreprintnumbers,     
               aps,                 
               prl,                 
               a4paper,             
               nofootinbib,         
               tightenlines,        
               floats,floatfix,      
]{revtex4-1}
\usepackage{amsmath,amsfonts,amssymb,bbm}
\usepackage[mathscr]{eucal}
\usepackage{tensor}
\usepackage[T1]{fontenc}
\usepackage[utf8]{inputenc}
\usepackage{lmodern}
\usepackage{color}
\usepackage{collref}
\usepackage{hyperref}
\usepackage{graphicx}
\usepackage{float}
\usepackage{color}
\usepackage{tikz}
\usetikzlibrary{calc,decorations.markings}
\allowdisplaybreaks[1]
\hypersetup{
colorlinks=true,
linktoc=page,    
linkcolor=blue,
citecolor=blue,
urlcolor=blue,
colorlinks=true,
}

\newcommand\td{\text{d}}
\newcommand\cO{{\cal O}}
\newcommand{\p}{\partial}

\newcommand{\be}{\begin{equation}}
\newcommand{\ee}{\end{equation}}
\newcommand{\bea}{\begin{eqnarray}}
\newcommand{\eea}{\end{eqnarray}}

\def\bz{\bar z}

\newcommand*\xbar[1]{%
  \hbox{%
    \vbox{%
      \hrule height 0.5pt 
      \kern0.3ex
      \hbox{%
        \kern-0.0em
        \ensuremath{#1}%
        \kern-0.0em
      }%
    }%
  }%
}
\def \red{{}}
\def \blue{{}}

\begin{document}

\title{\red{Supertranslations in the bulk of spacetime}}

\author{Pujian Mao}
\email{pjmao@tju.edu.cn}

\affiliation{
Center for Joint Quantum Studies, Department of Physics, School of Science, Tianjin University, 135 Yaguan Road, Tianjin 300350, China\\
}

\pacs{}

\begin{abstract}

Supertranslations are usually defined as asymptotic symmetries associated with spacetime boundaries, such as null infinity and black hole horizons. In this Letter, we show that supertranslations admit a natural, coordinate-independent extension into the bulk of spacetime, realized as transitions between families of null hypersurfaces. This construction applies to generic spacetimes \blue{admitting null boundaries with residual symmetries} and unifies the realizations of supertranslations at null infinity and \blue{finite-distance null hypersurfaces such as black hole horizons}. The bulk supertranslation is connected to boundary supertranslation by characteristic flows. The associated symmetry algebra \red{at the linearized level} can be realized by light-ray operators defined on the null hypersurface \blue{and the bulk supertranslation acts as a zero-mode operator in the context of light-cone quantization}. Within this framework, the gravitational wave memory effect corresponds to a shift of null hypersurfaces in the bulk. As explicit examples, we compute bulk supertranslations in Minkowski spacetime and four-dimensional Schwarzschild spacetime, where we uncover a novel curvature-\red{dependent} memory effect with observable consequences for light propagation.

\end{abstract}

\maketitle

\textit{Introduction}. Supertranslation is a very surprising result from the investigation on asymptotic structure of the spacetime at null infinity \cite{Bondi:1962px,Sachs:1962wk,Sachs:1962zza}, where the spacetime symmetry is enhanced from the Poincar\'{e} group to the Bondi-Metzner-Sachs (BMS) group. The BMS group is the semidirect product of the Lorentz group and an infinite-dimensional Abelian group, the supertranslation that generalizes the ordinary translations. Although supertranslation symmetry was discovered long ago, its physical implications have only been appreciated recently \cite{strominger:2017zoo}. The major breakthrough was made by Strominger for the discovery of the BMS invariance of gravitational scattering \cite{Strominger:2013jfa}. Supertranslations lie at the heart of a fascinating triangular equivalence \cite{strominger:2017zoo}. It reveals the symmetry origin of the Weinberg’s soft graviton theorem at the quantum level  \cite{He:2014laa} and characterizes the vacua transition associated with the gravitational memory effect at the classical level \cite{Strominger:2014pwa}. Further developments from various perspectives can be found, e.g., in \cite{Carlip:2017xne,Bousso:2017rsx,Sousa:2017auc,Choi:2017ylo,Cordova:2018ygx,Donnay:2020fof,Chen:2021szm,Fuentealba:2021xhn,Chakraborty:2021sbc,Fuentealba:2021yvo,Veneziano:2022zwh,Javadinezhad:2022ldc,Fuentealba:2023syb,Henneaux:2023neb,Javadinezhad:2023mtp,Elkhidir:2024izo,DeLuca:2024cjl,DeLuca:2024asq,Henneaux:2025ocw}.

A natural question is whether supertranslations can arise in the bulk of spacetime \cite{Compere:2016jwb,Compere:2016hzt,Compere:2016gwf,CL}. Important progress was achieved with the discovery of near-horizon symmetries \cite{Donnay:2015abr,Donnay:2016ejv}, but their emergence relies on inner boundaries such as black hole horizons, which share structural features with null infinity\cite{Ashtekar:2024mme,Ashtekar:2024bpi,Ashtekar:2024stm}. A bulk supertranslation in generic spacetime has not yet been fully established \cite{early}, which is essential as bulk supertranslations could unify near-horizon and null-infinity symmetries and yield complete conservation laws. They are fundamental in the application of soft hairs in resolving the black hole information paradox \cite{Hawking:2015qqa,Hawking:2016msc} and in the description of vacua transition associated with the bulk gravitational memory \cite{Bart:2019gnf}. This also raises the intriguing possibility that supertranslations are not merely asymptotic or near horizon symmetries but encode physical effects throughout the bulk of spacetime.

Supertranslations were originally discovered with respect to particular gauge conditions in a specific coordinate system, such as the Bondi gauge \cite{Bondi:1962px,Sachs:1962wk} or the Newman-Unti (NU) gauge \cite{Newman:1961qr,Newman:1962cia} in the Newman-Penrose formalism \cite{Newman:1961qr}. Despite these differences, the gauges share the same asymptotic structure at null infinity \cite{Barnich:2011ty,Geiller:2022vto,Geiller:2024amx}, and supertranslations can be defined using purely geometric and conformal method \cite{Penrose:1962ij,Geroch:1977big,Wald:1984rg,Ashtekar:2014zsa}. Therefore, a proper definition of a bulk supertranslation requires not only specifying the physical structure it preserves but also formulating it in a coordinate-independent manner.

\blue{
In this Letter, we demonstrate that boundary supertranslations can induce transitions between null hypersurfaces in the bulk of spacetime, which are generated by characteristic flows. Hence, such transitions of null hypersurfaces naturally represent the bulk supertranslations. Since null hypersurfaces are geometric entities, this definition is inherently coordinate-independent. The construction does not rely on the existence of null infinity supertranslations. Any null hypersurface admitting a residual symmetry can serve as the starting point of the characteristic flow that generates the corresponding bulk supertranslation.
} 

Inspired by the intrinsic connection between supertranslation and gravitational memory \cite{Strominger:2014pwa}, we propose a bulk memory effect corresponding to the transition of null geodesics from one hypersurface to another. \blue{Thus, the bulk supertranslation defines a natural extension of boundary supertranslation symmetry into the bulk} with observable memory effects \cite{Goncharov:2023woe}. As concrete illustrations, we compute supertranslations for Minkowski spacetime in four dimensions, which recovers the bulk extension of supertranslations in Minkowski spacetime obtained in \cite{Compere:2016jwb}. \blue{The corresponding memory effect is encoded in the permanent change of the expansion and shear from the geodesic deviation.} The computations can be extended to Minkowski spacetime in higher dimensions. 

A novel curvature-dependent memory effect is uncovered from the supertranslation in four-dimensional Schwarzschild spacetime. \blue{Specifically, a light ray initially directed toward the black hole along a geodesic without turning points can be shifted to a new trajectory with a turning point by gravitational waves with memory, which deflects the light ray and allows it to return to the same side of the black hole. The existence of a turning point is a diffeomorphism-invariant property of a geodesic. Hence, the bulk memory yields a transition between distinct classes of null geodesics.} This type of memory effect occurs only in curved spacetime \cite{flat}, namely black hole memory \cite{Donnay:2018ckb,Rahman:2019bmk}, and can be directly determined within black hole perturbation theory \cite{Elhashash:2024thm,Cunningham:2024dog}.

\blue{The bulk supertranslation algebra \red{at the linearized level} can be realized by light-ray operators defined from light-ray integral of the energy-momentum tensor} \blue{on the null hypersurface. The bulk supertranslation acts as a zero-mode operator in the context of light-cone quantization \cite{Brodsky:1997de,Heinzl:2000ht}, which fully aligns with the spontaneously broken nature of the null infinity supertranslation \cite{strominger:2017zoo}.}


\textit{Residual gauge transformation and supertranslation}. In this section, we first examine the physical implications of residual gauge transformations within the NU gauge and identify the bulk supertranslations in this context. \blue{The construction of NU gauge is detailed in End Matter.} We then generalize these definitions to a generic coordinate system. Suppose that two NU coordinate systems $(u,r,x^a)$ and $(\bar u, \bar r, \bar x^a)$ are related by $\bar x^\mu=\bar x^\mu (x^\alpha)$. A key feature of the NU gauge is the choice of a null hypersurface labeled by a single coordinate. If the spacetime can be described in both NU coordinate systems, then the hypersurfaces $u=$const and $\bar u=$const are both null. Consequently, the full residual gauge transformations of the NU gauge are characterized by different families of null hypersurfaces \cite{hypersurface}. In NU coordinates $(u,r,x^a)$, a residual gauge transformation is determined by a scalar potential $\bar u$, with the condition that $\bar\ell=\td \bar u$ is a null vector. 

The physical significance of a residual gauge transformation lies in the change of null hypersurfaces. Consequently, the system can be formulated as a characteristic initial value problem based on different families of null hypersurfaces \cite{Winicour:2001imp,Chesler:2013lia}. The coordinate change itself is not fundamental, as it merely reflects a different parametrization of the null hypersurfaces. \blue{Nevertheless, the new coordinates manifests the characterization of the new null hypersurface. From an invariance  perspective, all residual gauge transformations preserve the foliation of the spacetime into null hypersurface.}

\blue{ 
Null foliations of spacetime play an important role in quantum field theory, particularly in light-cone quantization \cite{Brodsky:1997de,Heinzl:2000ht}. In such framework the choice of light-cone time corresponds to selecting a null hypersurface on which the Hamiltonian evolution is defined. Supertranslations naturally capture the freedom in the selection of null hypersurfaces. As we show below, supertranslations relating nearby hypersurfaces generate zero-mode operators constructed from the light-ray integral of energy–momentum tensor} \red{at the linearized level for perturbative theory on fixed spacetime background}. \blue{These operators act on the physical states of the theory and therefore represent symmetry transformations rather than coordinate redundancies. At the classical level, such transition of null hypersurfaces corresponds to observational memory effect.
}

Since $\bar\ell=\td \bar u$ is a null vector, any rescaling $\Phi \bar\ell$ remains null. This rescaling freedom can be used to write the null vector in the form $\bar \ell=\td \bar u =\td u + \td f(r,x^a)$, so that $\bar u=u+f(r,x^a)$. Imposing the null condition then leads to the constraint for $f$ as
\be\label{constraint}
g^{rr}(\p_r f)^2 - 2 \p_r f + 2 g^{r a} \p_r f \p_a f + g^{ab} \p_a f \p_b f=0.
\ee
This equation is precisely the condition $g^{\bar u\bar u}=0$ in the transformed NU coordinates. So any solution of \eqref{constraint} corresponds to a supertranslation arising from a finite BMS transformation \cite{Bondi:1962px,Sachs:1962wk,Sachs:1962zza,Barnich:2016lyg,Flanagan:2023jio}. In particular, the constraint \eqref{constraint} reduces to an ordinary differential equation for variable $r$ when considering a series expansion near the null infinity for asymptotically flat spacetime. The integration constant in $f$ at $\cO(1)$ in a $1/r$ expansion reproduces the standard BMS supertranslation at null infinity \cite{Bondi:1962px,Sachs:1962wk,Sachs:1962zza}. All the subleading terms in $f$ are uniquely fixed by \eqref{constraint}. Therefore, a supertranslation in the NU gauge throughout the bulk spacetime can be defined as any scalar function $f(r,x^a)$ such that $\td (u +  f)$ remains null.

A few immediate remarks regarding the generic definition of supertranslation in the bulk are as follows:
\begin{itemize}
    \item We have introduced a rescaling $\Phi$ to specify a supertranslation. However, this rescaling has its own significance, as it relates to various generalizations of BMS symmetry, including superrotations and Weyl-BMS symmetries, see, e.g., \cite{Barnich:2009se,Barnich:2010eb,Barnich:2010ojg,Barnich:2011mi,Barnich:2016lyg,Compere:2018ylh,Campiglia:2015yka,Campiglia:2020qvc,Flanagan:2015pxa,Flanagan:2023jio,Freidel:2021fxf}.
    \item The NU gauge was originally formulated in four dimensions, but it can be generalized to arbitrary dimensions. In our construction, the essential step is the selection of a family of null hypersurfaces. Consequently, the residual gauge transformations and supertranslations described above can be straightforwardly extended to higher dimensions.
    \item The definition of supertranslation does not depend on a specific spacetime boundary. For instance, inserting solutions in a series expansion with a cosmological constant, the resulting supertranslations are consistent with the $\Lambda$-BMS transformations derived in \cite{Compere:2019bua,Compere:2020lrt}.
    \item Applying a near-horizon series expansion, the bulk supertranslation reduces to the near-horizon supertranslations identified in \cite{Donnay:2015abr,Donnay:2016ejv}, and also \cite{Afshar:2016wfy,Shi:2016jtn,Akhmedov:2017ftb,Chandrasekaran:2018aop,Haco:2018ske,Grumiller:2019fmp,Adami:2020uwd,Adami:2020amw,Chen:2020nyh,Adami:2020ugu,Chandrasekaran:2021hxc,Adami:2021nnf} for other realizations. The bulk supertranslation thus unifies the null infinity and near-horizon symmetries, providing a natural framework for matching conserved quantities at different boundaries, see, e.g., relevant investigations in \cite{Hawking:2016msc,Grumiller:2019ygj,Ruzziconi:2025fct,Agrawal:2025fsv,Ruzziconi:2025fuy}.
\end{itemize}

The investigations above are in NU coordinates. However, the key feature resides in the transition of null hypersurfaces. Therefore, one can define residual gauge transformations without reference to any particular coordinate system.

\textbf{Definition 1:} \textit{In a given coordinate system $x^\mu$, the residual gauge transformation is defined from any scalar function $H(x^\mu)$ such that $\td H(x^\mu)$ is a null vector.
}

Supertranslations are defined as the set of all residual gauge transformations modulo rescaling. In asymptotically flat spacetimes, they can be specified using a normalized timelike coordinate $t$ at infinity. A supertranslation can be defined as follows:  

\textbf{Definition 2:} \textit{In a given coordinate system $(t,x^i)$, a supertranslation is defined from any scalar function $h(x^i)$ such that $\td (t+h)$ is a null vector.
}

This definition uniquely specifies a supertranslation throughout the bulk of spacetime and is not restricted to the NU gauge. It can be applied directly to solutions that are difficult to express in NU coordinates, such as the Kerr solution \cite{Fletcher-Lun,Fletcher-Lun2003,Venter:2005cs}.

The scalar function specifying a supertranslation satisfies a nonlinear, first-order partial differential equation (PDE), which can be solved using the method of characteristic flows \cite{Evans,Levandosky}. In general, characteristic curves determine solutions implicitly. However, the new variables defined along the characteristics can be treated as new coordinates. In this way, the full coordinate transformation corresponding to a supertranslation can be explicitly constructed. In the following sections, we will derive supertranslations using characteristic flows for several exact solutions.


\textit{Four-dimensional Minkowski with plane boundary}. The Minkowski line-element with a plane boundary in flat null coordinates $(u,r,z,\bz)$ \cite{He:2019jjk} is given by
\be\label{plane}
\td s^2=- 2 \td u \td r  + 2 r^2 \td z \td\bz.
\ee
The null condition for the scalar function then reads
\be\label{f}
\p_{r} f=\frac{1}{r^2} \p_{z} f \p_{\bz} f.
\ee
This is a first-order fully nonlinear PDE, which can be solved by the method of characteristics following closely \cite{Levandosky}. The equation can be organized as
\be
F(r,z,\bz,p_r,p,\bar p)=p_r - \frac{1}{r^2} p \bar p=0,
\ee
where we introduce new variables $p_r=\p_r f$, $p=\p_{z} f$, and $\bar p=\p_{\bz} f$. The characteristic equations for this system are
\be
\begin{split}
&\frac{\td r}{\td s}=1,\quad     \frac{\td z}{\td s}=- \frac{\bar p}{r^2},  \quad \frac{\td \bz}{\td s}=- \frac{p}{r^2},\\
& \frac{\td p}{\td s}=0,\quad \frac{\td \bar p}{\td s}=0,\quad \frac{\td f}{\td s}= p \frac{\td z}{\td s} + \bar p \frac{\td \bz}{\td s} + \p_r f \frac{\td r}{\td s},
\end{split}
\ee
\blue{where $s$ parametrizes the characteristic curve.} Clearly, $p$ and $\bar p$ are constants along characteristics, which yields that 
\be\label{boundary}
z=\zeta+\frac{\bar p}{r},\qquad \bz=\bar \zeta+\frac{p}{r},
\ee
where $\zeta$ and $\bar\zeta$ are constants labeling different characteristic curves. Applying the differential equation \eqref{f}, one can obtain that
\be
\frac{\td f}{\td s}= - \frac{p \bar p}{r^2},\qquad f=f_0(\zeta,\bar \zeta)+\frac{p \bar p}{r}.
\ee
Since $p$ and $\bar p$ are constants on the characteristic curves, they can be fixed by the initial data at the infinity $r\to \infty$. Hence, $p =\p_{\zeta}f_0$ and $\bar p =\p_{\bar \zeta}f_0$. Thus, we have solved the equation \eqref{f} implicitly.

The solution for a bulk supertranslation $f$ is determined from the boundary variables $(\zeta,\bar\zeta)$ defined in \eqref{boundary}. This corresponds precisely to the characteristic flow of a boundary supertranslation $f_0$. The null hypersurface characterized by $f$ is therefore the characteristic flow of a cross-section at null infinity \cite{Geroch:1977big} specified by the boundary supertranslation $f_0$.

Considering $(\zeta,\bar \zeta)$ as new coordinates, one can obtain the following relation of the two coordinate systems
\be
u=\bar u-f_0-\frac{\p_\zeta f_0 \p_{\bar \zeta} f_0}{r},\qquad
z=\zeta+\frac{\p_{\bar\zeta} f_0}{r},
\ee
which precisely reproduces the transformation derived in \cite{Compere:2016jwb} for a supertranslated Minkowski spacetime \cite{compere}.

We now pinpoint the supertranslation effect from the new coordinates $(\bar u,\bar r,\zeta,\bar\zeta)$. Importantly, this is not merely the introduction of a new coordinate system, but rather the selection of a new family of null hypersurfaces with normal vector $\bar\ell=\td \bar u$. In other words, the objects of interest change from null geodesics on the hypersurfaces $u$=const to null geodesics on the hypersurfaces $\bar u$=const. Null geodesics lying on different hypersurfaces can exhibit distinct deviation properties, characterized by their expansion and shear. The two families of null hypersurfaces intersect null infinity at different cross-sections, which are related by a null infinity supertranslation.

From a dynamical perspective, supertranslations are equivalent to gravitational memory \cite{Strominger:2014pwa}. In the bulk, gravitational waves with memory will distort null geodesics \cite{Bart:2019gnf}, effectively forcing their trajectories to the transition from one null hypersurface to another. \blue{For the transition in Minkowski spacetime discussed above, the null congruence is initially set on the hypersurface $u$=const, with tangent vector $\ell=\td u$. Choosing a displacement vector $n=\td r$, the geodesic deviation of $\ell$ can be computed directly from the line-element in Eq. \eqref{plane}. In this configuration, the shear vanishes and the expansion is $-2/r$. }

\blue{After the passage of a gravitational wave with memory, the null congruence is mapped onto the hypersurface $\bar u$=const, with tangent vector $\bar\ell=\td \bar u$ and displacement vector $\bar n=\td \bar r$. The corresponding components of the shear tensor are
\be
\begin{split}
&\hat\sigma_{\zeta\zeta}=\frac{\p^2_\zeta f_0 \left[(\bar r + \p_\zeta \p_{\bar\zeta} f_0)^2 + \p^2_\zeta f_0 \p^2_{\bar\zeta} f_0 \right] } {(\bar r + \p_\zeta \p_{\bar\zeta} f_0)^2 - \p^2_\zeta f_0 \p^2_{\bar\zeta} f_0}, \\
&\hat\sigma_{\zeta\bar\zeta}=\frac{2 (\bar r + \p_\zeta \p_{\bar\zeta} f_0) \p^2_\zeta f_0 \p^2_{\bar\zeta} f_0  } {(\bar r + \p_\zeta \p_{\bar\zeta} f_0)^2 - \p^2_\zeta f_0 \p^2_{\bar\zeta} f_0},
\end{split}
\ee
and the expansion is
\be
\theta=-\frac{2 (\bar r + \p_\zeta \p_{\bar\zeta} f_0) } {(\bar r + \p_\zeta \p_{\bar\zeta} f_0)^2 - \p^2_\zeta f_0 \p^2_{\bar\zeta} f_0}.
\ee
In the Newman-Penrose formalism, the asymptotic shear of $\bar\ell$ is $\sigma_0=\p_{\bar\zeta}^2 f_0$, which reproduces the standard finite supertranslation at null infinity \cite{Barnich:2016lyg}. The permanent change of this asymptotic shear captures the gravitational memory at null infinity \cite{Frauendiener}. In this way, the bulk congruence deformation provides a simple reference for understanding how the null congruence evolves under supertranslations that is consistent with the well-known null infinity results.}

\begin{figure}[htbp]
    \centering
    \includegraphics[width=0.95\linewidth]{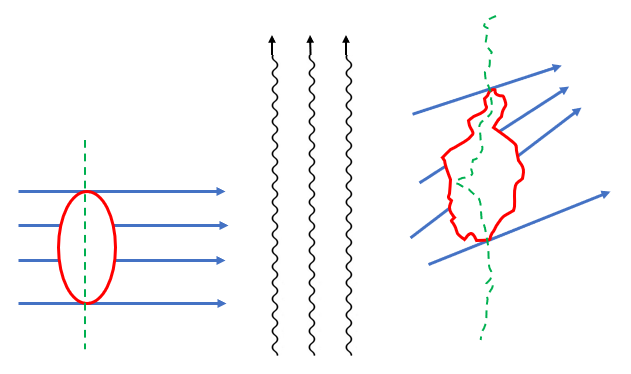}
    \caption{The black wavy lines represent gravitational waves with memory. The blue straight lines denote null geodesics. Initially, the null geodesics are plane-fronted (green dashed line) with vanishing shear. As the gravitational waves pass, they deform this congruence. The individual geodesic remains straight, but the associated light front is no longer planar. When described using the original geodesic parameter $r$, the light front is sampled at different values of $r$ from different rays, which manifests itself as a shear of the null congruence. }
    \label{memory}
\end{figure}

\blue{Concrete realizations of such memory can arise in impulsive gravitational waves, where null congruences experience a distributional jump in their expansion and shear across the impulse \cite{Steinbauer:1997dw,OLoughlin:2018ebk,Steinbauer:2018iis,Bhattacharjee:2019jaf}, as well as in soldering transformations across null hypersurfaces \cite{Blau:2015nee,Blau:2016juv,Bhattacharjee:2017gkh}, see e.g. the illustrations in Fig. \ref{memory}.}

\blue{The supertranslation in Minkowski spacetime in higher dimensions can be obtained in a similar way, see End Matter for the illustration with a sphere boundary.} Unlike Minkowski spacetime, where bulk supertranslations do not introduce qualitatively new structures, spacetime curvature leads to genuinely new features, which we now illustrate from the Schwarzschild solution.


\textit{Schwarzschild solution in four dimensions}. The line-element of Schwarzschild solution is given by
\be
\td s^2=-\left(1-\frac{2M}{r}\right)\td u^2 - 2 \td u \td r + r^2 \gamma_{ab}(x^c)\td x^a \td x^b.
\ee
The null condition for the scalar function is
\be
2\p_r f = \left(1-\frac{2M}{r}\right) (\p_r f)^2 + \frac{1}{r^2} \gamma^{ab}\p_a f \p_b f.
\ee
Solving this equation via the method of characteristic flows gives
\be\label{Schf}
f= f_0 ({\bar x}^a) + \int_{+\infty}^r \frac{\td r'}{ 1-\frac{2M}{r'} } \left( 1-  \frac{1}{\sqrt{1-\frac{L^2}{{r'}^2} \left(1-\frac{2M}{r'}\right) }} \right),
\ee
where $L^2={\bar\gamma}^{ab}\p_{{\bar x}^a} f_0 \p_{{\bar x}^b} f_0$. As before, $f_0$ represents the supertranslation at null infinity. This integral generally involves elliptic functions and can exhibit branch points depending on the supertranslation field $f_0$. These branch points are turning points of null geodesics on the associated null hypersurface, 
\be
\frac{\td r}{\td s} = - 2 \sqrt{1-\frac{L^2}{{r}^2} \left(1-\frac{2M}{r}\right) }=0.
\ee
Those branch points should correspond to points on the supertranslation horizon discovered in \cite{Compere:2016hzt}.

\blue{ 
In Minkowski spacetime, supertranslations distort null rays but do not generate turning points along their trajectories. In contrast, in Schwarzschild spacetime, a supertranslation can map null geodesics on a hypersurface that initially contains no turning points to a new hypersurface where turning points appear and are indicated by branch points of the supertranslation.
}

\blue{The existence of a turning point is a diffeomorphism-invariant property of a null geodesic. Hence, the bulk supertranslation represents a transition between distinct classes of null trajectories. From dynamical perspective, supertranslation corresponds to memory effect \cite{Strominger:2014pwa}. Gravitational waves with memory permanently change the class of null generators associated with a hypersurface. This provides a novel bulk realization of gravitational memory that does not rely on asymptotic boundaries. Because the effect arises within curved black hole backgrounds, it can be computed explicitly using black hole perturbation theory, where the new bulk memory effect arises as a gravitational lensing phenomenon induced by gravitational waves \cite{Labeyrie,Durrer:1994uu,Kaiser:1996wk,Damour:1998jm,Zhong:2024ysg}. This interpretation provides a concrete observational probe based on light propagation \cite{displacement}. For a fixed emitter–observer configuration, the passage of gravitational radiation induces a measurable shift in the reception position and arrival angle of null rays, associated with the transition between geodesics with and without turning points.
}

\begin{figure}[htbp]
    \centering
    \includegraphics[width=0.95\linewidth]{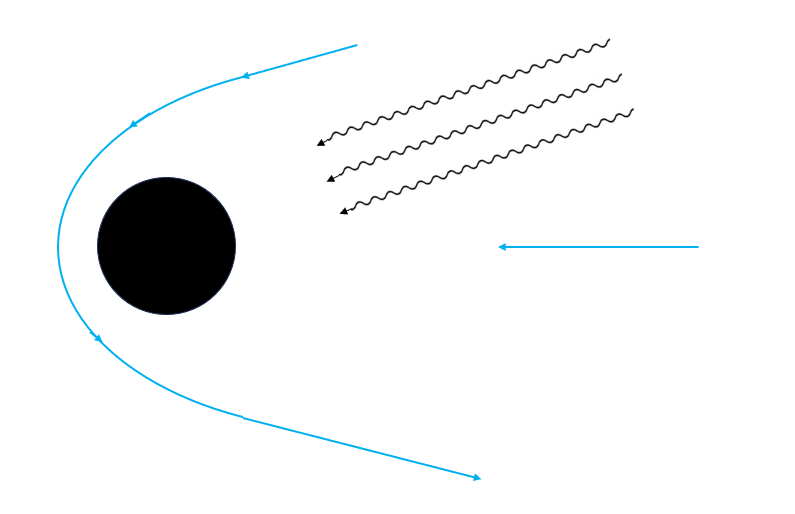}
    \caption{The black region represents the black hole. The black wavy lines indicate gravitational waves with memory, while the blue lines show the null geodesic before and after the memory effect. The memory effect can be observed from the propagation of the light ray.}
    \label{Schwarzschild}
\end{figure}

\red{
If the waveform of the gravitational radiation is known, the corresponding supertranslation $f_0$ can be obtained. For simplicity, considering the axi-symmetric case, where the waveform is $\phi$-independent, the trajectory of null geodesics is determined by
\be
\frac{\td \theta}{\td r}=-\frac{L}{r^2  \sqrt{1-\frac{L^2}{{r}^2} \left(1-\frac{2M}{r}\right) }},
\ee
where $L=\p_\theta f_0$.} \blue{In the absence of gravitational radiation $L=0$, the ray would cross the horizon and never return. However, as gravitational waves pass through the spacetime, they induce a nonzero angular momentum} \red{determined by the supertranslation $f_0$} \blue{for the light ray. As a result, the trajectory develops a turning point} \red{at $r^3-L^2 r + 2M L^2=0$} \blue{and then the light ray is deflected, eventually returning to the same side of the black hole where it can be detected by an observer. The qualitative behavior of this process is illustrated in Fig. \ref{Schwarzschild}.} \red{From the gravitational lensing perspective, the whole effect is induced by both the black hole and the gravitational waves, and it is indeed dominated by the black hole. Nevertheless, the initially light ray is specially arranged and cannot be deflected without gravitational waves. Hence, it is not necessary to separate or distinguish the two contributions. Receiving the light from the same side of black hole identifies the memory effect from gravitational waves.}

\blue{\textit{Supertranslation generators from light-ray operators}. Having established the geometric consequence of bulk supertranslations, we now discuss their realization as symmetry generators acting on fields defined on the null hypersurface in the context of light-cone quantization. The vacuum defined on a null hypersurface is trivial, and long-range (zero-momentum) phenomena are captured by the zero-mode sector, represented by zero-mode operators \cite{Brodsky:1997de,Heinzl:2000ht}. Supertranslations at null infinity are spontaneously broken and create Goldstone excitations belonging to the soft sector \cite{strominger:2017zoo}. In the bulk, a supertranslation corresponds to the transition from one null hypersurface to another, which precisely induces zero-mode quanta, as we will demonstrate.}

\blue{  
Consider two null hypersurfaces, $\Sigma$ and $\overline{\Sigma}$, related by a supertranslation, with null normals $\ell$ and $\bar{\ell}$ respectively. Using light-cone coordinates $(x^+,x^-,x_\bot)$ with $x^+$ as the evolution parameter specifying $\Sigma$, the Hamiltonian and longitudinal momentum are defined from the energy-momentum tensor as \cite{Brodsky:1997de,Heinzl:2000ht}
\be
P_+=\int_\Sigma \td x^- \td^2 x_\bot T^{+-},\quad
P_-=\int_\Sigma \td x^- \td^2 x_\bot T^{++}.
\ee
The vacuum on $\Sigma$ is annihilated by both $P_+$ and $P_-$, reflecting its simple structure. 
}

\blue{
The supertranslation is characterized by the new normal direction $\bar\ell=\td \bar x^+$. On $\overline{\Sigma}$, one can introduce the generator associated to $\bar\ell$ as (see End Matter for details)
\be
{\overline P}=\int_{\overline\Sigma} \sqrt{\bar q_{ij}}\, \td \bar x^- \td^2 \bar x_\bot  \overline T^{++},
\ee
which captures the supertranslation information. If the two hypersurfaces $\Sigma$ and $\overline{\Sigma}$ are infinitesimally separated, ${\overline P}$ can be pulled back to $\Sigma$, yielding to first order
\be
{\overline P}=P_- + 2\epsilon \int_\Sigma \td x^- \td^2 x_\bot  \p_i f(x_\bot) T^{+i} + \cO(\epsilon^2).
\ee
The leading-order correction term defines the supertranslation generator on $\Sigma$. This operator commutes with both $P_+$ and $P_-$ and therefore generates a symmetry transformation in the Hamiltonian framework. If it does not annihilate the vacuum, it acts as a zero-mode operator on $\Sigma$. This result perfectly reproduces the physical effects of null infinity supertranslations \cite{strominger:2017zoo} in the bulk. The commutator between two supertranslation generators appears only at $\epsilon^2$, so the algebra is Abelian at linear order. A more complete investigation of these operators will be presented elsewhere \cite{mao}.}

\blue{
Within the Hamiltonian framework, specifying an evolution direction breaks manifest diffeomorphism invariance, and generators are defined relative to the chosen null foliation. The construction is well defined in the semiclassical regime, for instance in linearized gravity on a fixed background, see, e.g., the Hamiltonian for linearized gravity in \cite{Bengtsson:1983pd,Ananth:2006fh}. The possible extension to full nonlinear gravity may require a quasi-local definition of energy–momentum tensor on null hypersurfaces, for example along the lines of the Brown–York construction \cite{Brown:1992br,Parattu:2016trq,Chandrasekaran:2021hxc}.} \red{Nevertheless, a fully diffeomorphism-invariant realization of the associated bulk generators in dynamical gravity remains an open problem, reflecting the broader challenge of defining gauge-invariant bulk observables in gravity \cite{Donnelly:2016rvo}.}


\blue{
\textit{Conclusions and perspectives}. In this Letter, we have introduced a coordinate-independent framework for bulk supertranslations based on characteristic flows on null hypersurfaces from boundary supertranslation. The bulk supertranslation modifies the null geodesic structure and leads to a novel} \red{curvature-dependent} \blue{memory effect. The light-ray operator representation suggests a natural realization of bulk supertranslation generators.
}

\blue{
A notable feature of our construction resides in the characteristic flow. In the context of AdS/CFT, the gravity dual of a deformation by the composite operator $T\bar T$ \cite{Smirnov:2016lqw,Cavaglia:2016oda} (also \cite{Jiang:2019epa,He:2025ppz} for reviews) corresponds to moving the boundary to a finite radial cutoff \cite{McGough:2016lol,Kraus:2018xrn,Hartman:2018tkw,Guica:2019nzm,Jafari:2019qns,He:2023hoj,He:2023knl,moving}. The associated $T\bar T$ flow has been identified as a characteristic flow \cite{Hou:2022csf}. These parallels suggest that our method of flowing asymptotic symmetries into the bulk is naturally compatible with the AdS/CFT perspective and merits further investigation.
}


\textit{Acknowledgments}. The author would like to thank Hamed Adami, Glenn Barnich, Luca Ciambelli, Jia Du, Shaoqi Hou, Delong Kong, Yue-Zhou Li, Shahin Sheikh-Jabbari, Vahid Taghiloo, Yu Tian, Hongbao Zhang, Ming Zhang, and Hua Xing Zhu for useful discussions. This work is supported in part by the National Natural Science Foundation of China (NSFC) under Grants No.~12475059 and No.~11935009, and by Tianjin University Self-Innovation Fund Extreme Basic Research Project Grant No. 2025XJ21-0007.


\begin{thebibliography}{84}%
\makeatletter
\providecommand \@ifxundefined [1]{%
 \@ifx{#1\undefined}
}%
\providecommand \@ifnum [1]{%
 \ifnum #1\expandafter \@firstoftwo
 \else \expandafter \@secondoftwo
 \fi
}%
\providecommand \@ifx [1]{%
 \ifx #1\expandafter \@firstoftwo
 \else \expandafter \@secondoftwo
 \fi
}%
\providecommand \natexlab [1]{#1}%
\providecommand \enquote  [1]{``#1''}%
\providecommand \bibnamefont  [1]{#1}%
\providecommand \bibfnamefont [1]{#1}%
\providecommand \citenamefont [1]{#1}%
\providecommand \href@noop [0]{\@secondoftwo}%
\providecommand \href [0]{\begingroup \@sanitize@url \@href}%
\providecommand \@href[1]{\@@startlink{#1}\@@href}%
\providecommand \@@href[1]{\endgroup#1\@@endlink}%
\providecommand \@sanitize@url [0]{\catcode `\\12\catcode `\$12\catcode
  `\&12\catcode `\#12\catcode `\^12\catcode `\_12\catcode `\%12\relax}%
\providecommand \@@startlink[1]{}%
\providecommand \@@endlink[0]{}%
\providecommand \url  [0]{\begingroup\@sanitize@url \@url }%
\providecommand \@url [1]{\endgroup\@href {#1}{\urlprefix }}%
\providecommand \urlprefix  [0]{URL }%
\providecommand \Eprint [0]{\href }%
\providecommand \doibase [0]{https://doi.org/}%
\providecommand \selectlanguage [0]{\@gobble}%
\providecommand \bibinfo  [0]{\@secondoftwo}%
\providecommand \bibfield  [0]{\@secondoftwo}%
\providecommand \translation [1]{[#1]}%
\providecommand \BibitemOpen [0]{}%
\providecommand \bibitemStop [0]{}%
\providecommand \bibitemNoStop [0]{.\EOS\space}%
\providecommand \EOS [0]{\spacefactor3000\relax}%
\providecommand \BibitemShut  [1]{\csname bibitem#1\endcsname}%
\let\auto@bib@innerbib\@empty
\bibitem [{\citenamefont {Bondi}\ \emph {et~al.}(1962)\citenamefont {Bondi},
  \citenamefont {van~der Burg},\ and\ \citenamefont {Metzner}}]{Bondi:1962px}%
  \BibitemOpen
  \bibfield  {author} {\bibinfo {author} {\bibfnamefont {H.}~\bibnamefont
  {Bondi}}, \bibinfo {author} {\bibfnamefont {M.~G.~J.}\ \bibnamefont {van~der
  Burg}},\ and\ \bibinfo {author} {\bibfnamefont {A.~W.~K.}\ \bibnamefont
  {Metzner}},\ }\bibfield  {title} {\bibinfo {title} {{Gravitational waves in
  general relativity. 7. Waves from axisymmetric isolated systems}},\ }\href
  {https://doi.org/10.1098/rspa.1962.0161} {\bibfield  {journal} {\bibinfo
  {journal} {Proc. Roy. Soc. Lond. A}\ }\textbf {\bibinfo {volume} {269}},\
  \bibinfo {pages} {21} (\bibinfo {year} {1962})}\BibitemShut {NoStop}%
\bibitem [{\citenamefont {Sachs}(1962{\natexlab{a}})}]{Sachs:1962wk}%
  \BibitemOpen
  \bibfield  {author} {\bibinfo {author} {\bibfnamefont {R.~K.}\ \bibnamefont
  {Sachs}},\ }\bibfield  {title} {\bibinfo {title} {{Gravitational waves in
  general relativity. 8. Waves in asymptotically flat space-times}},\ }\href
  {https://doi.org/10.1098/rspa.1962.0206} {\bibfield  {journal} {\bibinfo
  {journal} {Proc. Roy. Soc. Lond. A}\ }\textbf {\bibinfo {volume} {270}},\
  \bibinfo {pages} {103} (\bibinfo {year} {1962}{\natexlab{a}})}\BibitemShut
  {NoStop}%
\bibitem [{\citenamefont {Sachs}(1962{\natexlab{b}})}]{Sachs:1962zza}%
  \BibitemOpen
  \bibfield  {author} {\bibinfo {author} {\bibfnamefont {R.}~\bibnamefont
  {Sachs}},\ }\bibfield  {title} {\bibinfo {title} {{Asymptotic symmetries in
  gravitational theory}},\ }\href {https://doi.org/10.1103/PhysRev.128.2851}
  {\bibfield  {journal} {\bibinfo  {journal} {Phys. Rev.}\ }\textbf {\bibinfo
  {volume} {128}},\ \bibinfo {pages} {2851} (\bibinfo {year}
  {1962}{\natexlab{b}})}\BibitemShut {NoStop}%
\bibitem [{\citenamefont {Strominger}(2018)}]{strominger:2017zoo}%
  \BibitemOpen
  \bibfield  {author} {\bibinfo {author} {\bibfnamefont {A.}~\bibnamefont
  {Strominger}},\ }\href@noop {} {\emph {\bibinfo {title} {{Lectures on the
  Infrared Structure of Gravity and Gauge Theory}}}}\ (\bibinfo  {publisher}
  {Princeton University Press, Princeton},\ \bibinfo {year} {2018})\ \Eprint
  {https://arxiv.org/abs/1703.05448} {arXiv:1703.05448 [hep-th]} \BibitemShut
  {NoStop}%
\bibitem [{\citenamefont {Strominger}(2014)}]{Strominger:2013jfa}%
  \BibitemOpen
  \bibfield  {author} {\bibinfo {author} {\bibfnamefont {A.}~\bibnamefont
  {Strominger}},\ }\bibfield  {title} {\bibinfo {title} {{On BMS Invariance of
  Gravitational Scattering}},\ }\href {https://doi.org/10.1007/JHEP07(2014)152}
  {\bibfield  {journal} {\bibinfo  {journal} {JHEP}\ }\textbf {\bibinfo
  {volume} {07}},\ \bibinfo {pages} {152} (2014)},\ \Eprint
  {https://arxiv.org/abs/1312.2229} {arXiv:1312.2229 [hep-th]} \BibitemShut
  {NoStop}%
\bibitem [{\citenamefont {He}\ \emph {et~al.}(2015)\citenamefont {He},
  \citenamefont {Lysov}, \citenamefont {Mitra},\ and\ \citenamefont
  {Strominger}}]{He:2014laa}%
  \BibitemOpen
  \bibfield  {author} {\bibinfo {author} {\bibfnamefont {T.}~\bibnamefont
  {He}}, \bibinfo {author} {\bibfnamefont {V.}~\bibnamefont {Lysov}}, \bibinfo
  {author} {\bibfnamefont {P.}~\bibnamefont {Mitra}},\ and\ \bibinfo {author}
  {\bibfnamefont {A.}~\bibnamefont {Strominger}},\ }\bibfield  {title}
  {\bibinfo {title} {{BMS supertranslations and Weinberg{\textquoteright}s soft
  graviton theorem}},\ }\href {https://doi.org/10.1007/JHEP05(2015)151}
  {\bibfield  {journal} {\bibinfo  {journal} {JHEP}\ }\textbf {\bibinfo
  {volume} {05}},\ \bibinfo {pages} {151} (2015)},\ \Eprint
  {https://arxiv.org/abs/1401.7026} {arXiv:1401.7026 [hep-th]} \BibitemShut
  {NoStop}%
\bibitem [{\citenamefont {Strominger}\ and\ \citenamefont
  {Zhiboedov}(2016)}]{Strominger:2014pwa}%
  \BibitemOpen
  \bibfield  {author} {\bibinfo {author} {\bibfnamefont {A.}~\bibnamefont
  {Strominger}}\ and\ \bibinfo {author} {\bibfnamefont {A.}~\bibnamefont
  {Zhiboedov}},\ }\bibfield  {title} {\bibinfo {title} {{Gravitational Memory,
  BMS Supertranslations and Soft Theorems}},\ }\href
  {https://doi.org/10.1007/JHEP01(2016)086} {\bibfield  {journal} {\bibinfo
  {journal} {JHEP}\ }\textbf {\bibinfo {volume} {01}},\ \bibinfo {pages}
  {086} (2016)},\ \Eprint {https://arxiv.org/abs/1411.5745} {arXiv:1411.5745 [hep-th]}
  \BibitemShut {NoStop}%
\bibitem [{\citenamefont {Carlip}(2018)}]{Carlip:2017xne}%
  \BibitemOpen
  \bibfield  {author} {\bibinfo {author} {\bibfnamefont {S.}~\bibnamefont
  {Carlip}},\ }\bibfield  {title} {\bibinfo {title} {{Black Hole Entropy from
  Bondi-Metzner-Sachs Symmetry at the Horizon}},\ }\href
  {https://doi.org/10.1103/PhysRevLett.120.101301} {\bibfield  {journal}
  {\bibinfo  {journal} {Phys. Rev. Lett.}\ }\textbf {\bibinfo {volume} {120}},\
  \bibinfo {pages} {101301} (\bibinfo {year} {2018})},\ \Eprint
  {https://arxiv.org/abs/1702.04439} {arXiv:1702.04439 [gr-qc]} \BibitemShut
  {NoStop}%
\bibitem [{\citenamefont {Bousso}\ and\ \citenamefont
  {Porrati}(2017)}]{Bousso:2017rsx}%
  \BibitemOpen
  \bibfield  {author} {\bibinfo {author} {\bibfnamefont {R.}~\bibnamefont
  {Bousso}}\ and\ \bibinfo {author} {\bibfnamefont {M.}~\bibnamefont
  {Porrati}},\ }\bibfield  {title} {\bibinfo {title} {{Observable
  Supertranslations}},\ }\href {https://doi.org/10.1103/PhysRevD.96.086016}
  {\bibfield  {journal} {\bibinfo  {journal} {Phys. Rev. D}\ }\textbf {\bibinfo
  {volume} {96}},\ \bibinfo {pages} {086016} (\bibinfo {year} {2017})},\
  \Eprint {https://arxiv.org/abs/1706.09280} {arXiv:1706.09280 [hep-th]}
  \BibitemShut {NoStop}%
\bibitem [{\citenamefont {Sousa}\ \emph {et~al.}(2018)\citenamefont {Sousa},
  \citenamefont {Mil{\'a}ns~del Bosch},\ and\ \citenamefont
  {Reina}}]{Sousa:2017auc}%
  \BibitemOpen
  \bibfield  {author} {\bibinfo {author} {\bibfnamefont {K.}~\bibnamefont
  {Sousa}}, \bibinfo {author} {\bibfnamefont {G.}~\bibnamefont {Mil{\'a}ns~del
  Bosch}},\ and\ \bibinfo {author} {\bibfnamefont {B.}~\bibnamefont {Reina}},\
  }\bibfield  {title} {\bibinfo {title} {{Supertranslations: redundancies of
  horizon data, and global symmetries at null infinity}},\ }\href
  {https://doi.org/10.1088/1361-6382/aa9669} {\bibfield  {journal} {\bibinfo
  {journal} {Class. Quant. Grav.}\ }\textbf {\bibinfo {volume} {35}},\ \bibinfo
  {pages} {054002} (\bibinfo {year} {2018})},\ \Eprint
  {https://arxiv.org/abs/1707.02971} {arXiv:1707.02971 [hep-th]} \BibitemShut
  {NoStop}%
\bibitem [{\citenamefont {Choi}\ and\ \citenamefont
  {Akhoury}(2018)}]{Choi:2017ylo}%
  \BibitemOpen
  \bibfield  {author} {\bibinfo {author} {\bibfnamefont {S.}~\bibnamefont
  {Choi}}\ and\ \bibinfo {author} {\bibfnamefont {R.}~\bibnamefont {Akhoury}},\
  }\bibfield  {title} {\bibinfo {title} {{BMS Supertranslation Symmetry Implies
  Faddeev-Kulish Amplitudes}},\ }\href
  {https://doi.org/10.1007/JHEP02(2018)171} {\bibfield  {journal} {\bibinfo
  {journal} {JHEP}\ }\textbf {\bibinfo {volume} {02}},\ \bibinfo {pages}
  {171} (2018)},\ \Eprint {https://arxiv.org/abs/1712.04551} {arXiv:1712.04551
  [hep-th]} \BibitemShut {NoStop}%
\bibitem [{\citenamefont {C{\'o}rdova}\ and\ \citenamefont
  {Shao}(2018)}]{Cordova:2018ygx}%
  \BibitemOpen
  \bibfield  {author} {\bibinfo {author} {\bibfnamefont {C.}~\bibnamefont
  {C{\'o}rdova}}\ and\ \bibinfo {author} {\bibfnamefont {S.-H.}\ \bibnamefont
  {Shao}},\ }\bibfield  {title} {\bibinfo {title} {{Light-ray Operators and the
  BMS Algebra}},\ }\href {https://doi.org/10.1103/PhysRevD.98.125015}
  {\bibfield  {journal} {\bibinfo  {journal} {Phys. Rev. D}\ }\textbf {\bibinfo
  {volume} {98}},\ \bibinfo {pages} {125015} (\bibinfo {year} {2018})},\
  \Eprint {https://arxiv.org/abs/1810.05706} {arXiv:1810.05706 [hep-th]}
  \BibitemShut {NoStop}%
\bibitem [{\citenamefont {Donnay}\ \emph {et~al.}(2020)\citenamefont {Donnay},
  \citenamefont {Giribet},\ and\ \citenamefont {Rosso}}]{Donnay:2020fof}%
  \BibitemOpen
  \bibfield  {author} {\bibinfo {author} {\bibfnamefont {L.}~\bibnamefont
  {Donnay}}, \bibinfo {author} {\bibfnamefont {G.}~\bibnamefont {Giribet}},\
  and\ \bibinfo {author} {\bibfnamefont {F.}~\bibnamefont {Rosso}},\ }\bibfield
   {title} {\bibinfo {title} {{Quantum BMS transformations in conformally flat
  space-times and holography}},\ }\href
  {https://doi.org/10.1007/JHEP12(2020)102} {\bibfield  {journal} {\bibinfo
  {journal} {JHEP}\ }\textbf {\bibinfo {volume} {12}},\ \bibinfo {pages}
  {102} (2020)},\ \Eprint {https://arxiv.org/abs/2008.05483} {arXiv:2008.05483
  [hep-th]} \BibitemShut {NoStop}%
\bibitem [{\citenamefont {Chen}\ \emph {et~al.}(2021)\citenamefont {Chen},
  \citenamefont {Wang}, \citenamefont {Wang},\ and\ \citenamefont
  {Yau}}]{Chen:2021szm}%
  \BibitemOpen
  \bibfield  {author} {\bibinfo {author} {\bibfnamefont {P.-N.}\ \bibnamefont
  {Chen}}, \bibinfo {author} {\bibfnamefont {M.-T.}\ \bibnamefont {Wang}},
  \bibinfo {author} {\bibfnamefont {Y.-K.}\ \bibnamefont {Wang}},\ and\
  \bibinfo {author} {\bibfnamefont {S.-T.}\ \bibnamefont {Yau}},\ }\bibfield
  {title} {\bibinfo {title} {{Supertranslation invariance of angular
  momentum}},\ }\href {https://doi.org/10.4310/ATMP.2021.v25.n3.a4} {\bibfield
  {journal} {\bibinfo  {journal} {Adv. Theor. Math. Phys.}\ }\textbf {\bibinfo
  {volume} {25}},\ \bibinfo {pages} {777} (\bibinfo {year} {2021})},\ \Eprint
  {https://arxiv.org/abs/2102.03235} {arXiv:2102.03235 [gr-qc]} \BibitemShut
  {NoStop}%
\bibitem [{\citenamefont {Fuentealba}\ \emph {et~al.}(2021)\citenamefont
  {Fuentealba}, \citenamefont {Henneaux}, \citenamefont {Majumdar},
  \citenamefont {Matulich},\ and\ \citenamefont {Neogi}}]{Fuentealba:2021xhn}%
  \BibitemOpen
  \bibfield  {author} {\bibinfo {author} {\bibfnamefont {O.}~\bibnamefont
  {Fuentealba}}, \bibinfo {author} {\bibfnamefont {M.}~\bibnamefont
  {Henneaux}}, \bibinfo {author} {\bibfnamefont {S.}~\bibnamefont {Majumdar}},
  \bibinfo {author} {\bibfnamefont {J.}~\bibnamefont {Matulich}},\ and\
  \bibinfo {author} {\bibfnamefont {T.}~\bibnamefont {Neogi}},\ }\bibfield
  {title} {\bibinfo {title} {{Local supersymmetry and the square roots of
  Bondi-Metzner-Sachs supertranslations}},\ }\href
  {https://doi.org/10.1103/PhysRevD.104.L121702} {\bibfield  {journal}
  {\bibinfo  {journal} {Phys. Rev. D}\ }\textbf {\bibinfo {volume} {104}},\
  \bibinfo {pages} {L121702} (\bibinfo {year} {2021})},\ \Eprint
  {https://arxiv.org/abs/2108.07825} {arXiv:2108.07825 [hep-th]} \BibitemShut
  {NoStop}%
\bibitem [{\citenamefont {Chakraborty}\ \emph {et~al.}(2022)\citenamefont
  {Chakraborty}, \citenamefont {Ghosh}, \citenamefont {Hoque}, \citenamefont
  {Khairnar},\ and\ \citenamefont {Virmani}}]{Chakraborty:2021sbc}%
  \BibitemOpen
  \bibfield  {author} {\bibinfo {author} {\bibfnamefont {S.}~\bibnamefont
  {Chakraborty}}, \bibinfo {author} {\bibfnamefont {D.}~\bibnamefont {Ghosh}},
  \bibinfo {author} {\bibfnamefont {S.~J.}\ \bibnamefont {Hoque}}, \bibinfo
  {author} {\bibfnamefont {A.}~\bibnamefont {Khairnar}},\ and\ \bibinfo
  {author} {\bibfnamefont {A.}~\bibnamefont {Virmani}},\ }\bibfield  {title}
  {\bibinfo {title} {{Supertranslations at timelike infinity}},\ }\href
  {https://doi.org/10.1007/JHEP02(2022)022} {\bibfield  {journal} {\bibinfo
  {journal} {JHEP}\ }\textbf {\bibinfo {volume} {02}},\ \bibinfo {pages}
  {022} (2022)},\ \Eprint {https://arxiv.org/abs/2111.08907} {arXiv:2111.08907
  [hep-th]} \BibitemShut {NoStop}%
\bibitem [{\citenamefont {Fuentealba}\ \emph {et~al.}(2022)\citenamefont
  {Fuentealba}, \citenamefont {Henneaux}, \citenamefont {Matulich},\ and\
  \citenamefont {Troessaert}}]{Fuentealba:2021yvo}%
  \BibitemOpen
  \bibfield  {author} {\bibinfo {author} {\bibfnamefont {O.}~\bibnamefont
  {Fuentealba}}, \bibinfo {author} {\bibfnamefont {M.}~\bibnamefont
  {Henneaux}}, \bibinfo {author} {\bibfnamefont {J.}~\bibnamefont {Matulich}},\
  and\ \bibinfo {author} {\bibfnamefont {C.}~\bibnamefont {Troessaert}},\
  }\bibfield  {title} {\bibinfo {title} {{Bondi-Metzner-Sachs Group in Five
  Spacetime Dimensions}},\ }\href
  {https://doi.org/10.1103/PhysRevLett.128.051103} {\bibfield  {journal}
  {\bibinfo  {journal} {Phys. Rev. Lett.}\ }\textbf {\bibinfo {volume} {128}},\
  \bibinfo {pages} {051103} (\bibinfo {year} {2022})},\ \Eprint
  {https://arxiv.org/abs/2111.09664} {arXiv:2111.09664 [hep-th]} \BibitemShut
  {NoStop}%
\bibitem [{\citenamefont {Veneziano}\ and\ \citenamefont
  {Vilkovisky}(2022)}]{Veneziano:2022zwh}%
  \BibitemOpen
  \bibfield  {author} {\bibinfo {author} {\bibfnamefont {G.}~\bibnamefont
  {Veneziano}}\ and\ \bibinfo {author} {\bibfnamefont {G.~A.}\ \bibnamefont
  {Vilkovisky}},\ }\bibfield  {title} {\bibinfo {title} {{Angular momentum loss
  in gravitational scattering, radiation reaction, and the Bondi gauge
  ambiguity}},\ }\href {https://doi.org/10.1016/j.physletb.2022.137419}
  {\bibfield  {journal} {\bibinfo  {journal} {Phys. Lett. B}\ }\textbf
  {\bibinfo {volume} {834}},\ \bibinfo {pages} {137419} (\bibinfo {year}
  {2022})},\ \Eprint {https://arxiv.org/abs/2201.11607} {arXiv:2201.11607
  [gr-qc]} \BibitemShut {NoStop}%
\bibitem [{\citenamefont {Javadinezhad}\ and\ \citenamefont
  {Porrati}(2023)}]{Javadinezhad:2022ldc}%
  \BibitemOpen
  \bibfield  {author} {\bibinfo {author} {\bibfnamefont {R.}~\bibnamefont
  {Javadinezhad}}\ and\ \bibinfo {author} {\bibfnamefont {M.}~\bibnamefont
  {Porrati}},\ }\bibfield  {title} {\bibinfo {title}
  {{Supertranslation-Invariant Formula for the Angular Momentum Flux in
  Gravitational Scattering}},\ }\href
  {https://doi.org/10.1103/PhysRevLett.130.011401} {\bibfield  {journal}
  {\bibinfo  {journal} {Phys. Rev. Lett.}\ }\textbf {\bibinfo {volume} {130}},\
  \bibinfo {pages} {011401} (\bibinfo {year} {2023})},\ \Eprint
  {https://arxiv.org/abs/2211.06538} {arXiv:2211.06538 [gr-qc]} \BibitemShut
  {NoStop}%
\bibitem [{\citenamefont {Fuentealba}\ \emph {et~al.}(2023)\citenamefont
  {Fuentealba}, \citenamefont {Henneaux},\ and\ \citenamefont
  {Troessaert}}]{Fuentealba:2023syb}%
  \BibitemOpen
  \bibfield  {author} {\bibinfo {author} {\bibfnamefont {O.}~\bibnamefont
  {Fuentealba}}, \bibinfo {author} {\bibfnamefont {M.}~\bibnamefont
  {Henneaux}},\ and\ \bibinfo {author} {\bibfnamefont {C.}~\bibnamefont
  {Troessaert}},\ }\bibfield  {title} {\bibinfo {title} {{Asymptotic Symmetry
  Algebra of Einstein Gravity and Lorentz Generators}},\ }\href
  {https://doi.org/10.1103/PhysRevLett.131.111402} {\bibfield  {journal}
  {\bibinfo  {journal} {Phys. Rev. Lett.}\ }\textbf {\bibinfo {volume} {131}},\
  \bibinfo {pages} {111402} (\bibinfo {year} {2023})},\ \Eprint
  {https://arxiv.org/abs/2305.05436} {arXiv:2305.05436 [hep-th]} \BibitemShut
  {NoStop}%
\bibitem [{\citenamefont {Henneaux}(2024)}]{Henneaux:2023neb}%
  \BibitemOpen
  \bibfield  {author} {\bibinfo {author} {\bibfnamefont {M.}~\bibnamefont
  {Henneaux}},\ }\bibfield  {title} {\bibinfo {title}
  {{Corvino{\textendash}Schoen theorem and supertranslations at spatial
  infinity}},\ }\href {https://doi.org/10.1142/S0217751X24470079} {\bibfield
  {journal} {\bibinfo  {journal} {Int. J. Mod. Phys. A}\ }\textbf {\bibinfo
  {volume} {39}},\ \bibinfo {pages} {2447007} (\bibinfo {year} {2024})},\
  \Eprint {https://arxiv.org/abs/2306.12505} {arXiv:2306.12505 [gr-qc]}
  \BibitemShut {NoStop}%
\bibitem [{\citenamefont {Javadinezhad}\ and\ \citenamefont
  {Porrati}(2024)}]{Javadinezhad:2023mtp}%
  \BibitemOpen
  \bibfield  {author} {\bibinfo {author} {\bibfnamefont {R.}~\bibnamefont
  {Javadinezhad}}\ and\ \bibinfo {author} {\bibfnamefont {M.}~\bibnamefont
  {Porrati}},\ }\bibfield  {title} {\bibinfo {title} {{Three Puzzles with
  Covariance and Supertranslation Invariance of Angular Momentum Flux and Their
  Solutions}},\ }\href {https://doi.org/10.1103/PhysRevLett.132.151604}
  {\bibfield  {journal} {\bibinfo  {journal} {Phys. Rev. Lett.}\ }\textbf
  {\bibinfo {volume} {132}},\ \bibinfo {pages} {151604} (\bibinfo {year}
  {2024})},\ \Eprint {https://arxiv.org/abs/2312.02458} {arXiv:2312.02458
  [hep-th]} \BibitemShut {NoStop}%
\bibitem [{\citenamefont {Elkhidir}\ \emph {et~al.}(2025)\citenamefont
  {Elkhidir}, \citenamefont {O'Connell},\ and\ \citenamefont
  {Roiban}}]{Elkhidir:2024izo}%
  \BibitemOpen
  \bibfield  {author} {\bibinfo {author} {\bibfnamefont {A.}~\bibnamefont
  {Elkhidir}}, \bibinfo {author} {\bibfnamefont {D.}~\bibnamefont
  {O'Connell}},\ and\ \bibinfo {author} {\bibfnamefont {R.}~\bibnamefont
  {Roiban}},\ }\bibfield  {title} {\bibinfo {title} {{Supertranslations from
  Scattering Amplitudes}},\ }\href {https://doi.org/10.1103/fxdk-5qwc}
  {\bibfield  {journal} {\bibinfo  {journal} {Phys. Rev. Lett.}\ }\textbf
  {\bibinfo {volume} {135}},\ \bibinfo {pages} {151601} (\bibinfo {year}
  {2025})},\ \Eprint {https://arxiv.org/abs/2408.15961} {arXiv:2408.15961
  [hep-th]} \BibitemShut {NoStop}%
\bibitem [{\citenamefont {De~Luca}\ \emph
  {et~al.}(2025{\natexlab{a}})\citenamefont {De~Luca}, \citenamefont {Khoury},\
  and\ \citenamefont {Wong}}]{DeLuca:2024cjl}%
  \BibitemOpen
  \bibfield  {author} {\bibinfo {author} {\bibfnamefont {V.}~\bibnamefont
  {De~Luca}}, \bibinfo {author} {\bibfnamefont {J.}~\bibnamefont {Khoury}},\
  and\ \bibinfo {author} {\bibfnamefont {S.~S.~C.}\ \bibnamefont {Wong}},\
  }\bibfield  {title} {\bibinfo {title} {{Gravitational memory and soft
  theorems: The local perspective}},\ }\href
  {https://doi.org/10.1103/gbg1-mz49} {\bibfield  {journal} {\bibinfo
  {journal} {Phys. Rev. D}\ }\textbf {\bibinfo {volume} {112}},\ \bibinfo
  {pages} {L021502} (\bibinfo {year} {2025}{\natexlab{a}})},\ \Eprint
  {https://arxiv.org/abs/2412.01910} {arXiv:2412.01910 [gr-qc]} \BibitemShut
  {NoStop}%
\bibitem [{\citenamefont {De~Luca}\ \emph
  {et~al.}(2025{\natexlab{b}})\citenamefont {De~Luca}, \citenamefont {Khoury},\
  and\ \citenamefont {Wong}}]{DeLuca:2024asq}%
  \BibitemOpen
  \bibfield  {author} {\bibinfo {author} {\bibfnamefont {V.}~\bibnamefont
  {De~Luca}}, \bibinfo {author} {\bibfnamefont {J.}~\bibnamefont {Khoury}},\
  and\ \bibinfo {author} {\bibfnamefont {S.~S.~C.}\ \bibnamefont {Wong}},\
  }\bibfield  {title} {\bibinfo {title} {{Gravitational memory and Ward
  identities in the local detector frame}},\ }\href
  {https://doi.org/10.1103/PhysRevD.112.024032} {\bibfield  {journal} {\bibinfo
   {journal} {Phys. Rev. D}\ }\textbf {\bibinfo {volume} {112}},\ \bibinfo
  {pages} {024032} (\bibinfo {year} {2025}{\natexlab{b}})},\ \Eprint
  {https://arxiv.org/abs/2412.12273} {arXiv:2412.12273 [gr-qc]} \BibitemShut
  {NoStop}%
\bibitem [{\citenamefont {Henneaux}(2025)}]{Henneaux:2025ocw}%
  \BibitemOpen
  \bibfield  {author} {\bibinfo {author} {\bibfnamefont {M.}~\bibnamefont
  {Henneaux}},\ }\bibfield  {title} {\bibinfo {title} {{Wheeler-DeWitt Equation
  and Bondi-Metzner-Sachs (BMS) Symmetry}},\ }\href
  {https://doi.org/10.1103/29w3-3mmc} {\bibfield  {journal} {\bibinfo
  {journal} {Phys. Rev. Lett.}\ }\textbf {\bibinfo {volume} {135}},\ \bibinfo
  {pages} {061501} (\bibinfo {year} {2025})},\ \Eprint
  {https://arxiv.org/abs/2506.02240} {arXiv:2506.02240 [hep-th]} \BibitemShut
  {NoStop}%
\bibitem [{\citenamefont {Comp{\`e}re}\ and\ \citenamefont
  {Long}(2016{\natexlab{a}})}]{Compere:2016jwb}%
  \BibitemOpen
  \bibfield  {author} {\bibinfo {author} {\bibfnamefont {G.}~\bibnamefont
  {Comp{\`e}re}}\ and\ \bibinfo {author} {\bibfnamefont {J.}~\bibnamefont
  {Long}},\ }\bibfield  {title} {\bibinfo {title} {{Vacua of the gravitational
  field}},\ }\href {https://doi.org/10.1007/JHEP07(2016)137} {\bibfield
  {journal} {\bibinfo  {journal} {JHEP}\ }\textbf {\bibinfo {volume} {07}},\
  \bibinfo {pages} {137} (2016)},\ \Eprint {https://arxiv.org/abs/1601.04958}
  {arXiv:1601.04958 [hep-th]} \BibitemShut {NoStop}%
\bibitem [{\citenamefont {Comp{\`e}re}\ and\ \citenamefont
  {Long}(2016{\natexlab{b}})}]{Compere:2016hzt}%
  \BibitemOpen
  \bibfield  {author} {\bibinfo {author} {\bibfnamefont {G.}~\bibnamefont
  {Comp{\`e}re}}\ and\ \bibinfo {author} {\bibfnamefont {J.}~\bibnamefont
  {Long}},\ }\bibfield  {title} {\bibinfo {title} {{Classical static final
  state of collapse with supertranslation memory}},\ }\href
  {https://doi.org/10.1088/0264-9381/33/19/195001} {\bibfield  {journal}
  {\bibinfo  {journal} {Class. Quant. Grav.}\ }\textbf {\bibinfo {volume}
  {33}},\ \bibinfo {pages} {195001} (\bibinfo {year} {2016}{\natexlab{b}})},\
  \Eprint {https://arxiv.org/abs/1602.05197} {arXiv:1602.05197 [gr-qc]}
  \BibitemShut {NoStop}%
\bibitem [{\citenamefont {Comp{\`e}re}(2016)}]{Compere:2016gwf}%
  \BibitemOpen
  \bibfield  {author} {\bibinfo {author} {\bibfnamefont {G.}~\bibnamefont
  {Comp{\`e}re}},\ }\bibfield  {title} {\bibinfo {title} {{Bulk
  supertranslation memories: a concept reshaping the vacua and black holes of
  general relativity}},\ }\href {https://doi.org/10.1142/S0218271816440065}
  {\bibfield  {journal} {\bibinfo  {journal} {Int. J. Mod. Phys. D}\ }\textbf
  {\bibinfo {volume} {25}},\ \bibinfo {pages} {1644006} (\bibinfo {year}
  {2016})},\ \Eprint {https://arxiv.org/abs/1606.00377} {arXiv:1606.00377
  [hep-th]} \BibitemShut {NoStop}%
\bibitem [{CL()}]{CL}%
  \BibitemOpen
  \href@noop {} {}\bibinfo {note} {The supertranslation on particular exact
  solutions has been studied in
  \cite{Compere:2016jwb,Compere:2016hzt,Compere:2016gwf}. However, those works
  mainly specify a bulk supertranslation from the asymptotic behavior. The
  critical issue as a symmetry, what it preserves in the bulk, is still
  unclear.}\BibitemShut {Stop}%
\bibitem [{\citenamefont {Donnay}\ \emph
  {et~al.}(2016{\natexlab{a}})\citenamefont {Donnay}, \citenamefont {Giribet},
  \citenamefont {Gonzalez},\ and\ \citenamefont {Pino}}]{Donnay:2015abr}%
  \BibitemOpen
  \bibfield  {author} {\bibinfo {author} {\bibfnamefont {L.}~\bibnamefont
  {Donnay}}, \bibinfo {author} {\bibfnamefont {G.}~\bibnamefont {Giribet}},
  \bibinfo {author} {\bibfnamefont {H.~A.}\ \bibnamefont {Gonzalez}},\ and\
  \bibinfo {author} {\bibfnamefont {M.}~\bibnamefont {Pino}},\ }\bibfield
  {title} {\bibinfo {title} {{Supertranslations and Superrotations at the Black
  Hole Horizon}},\ }\href {https://doi.org/10.1103/PhysRevLett.116.091101}
  {\bibfield  {journal} {\bibinfo  {journal} {Phys. Rev. Lett.}\ }\textbf
  {\bibinfo {volume} {116}},\ \bibinfo {pages} {091101} (\bibinfo {year}
  {2016}{\natexlab{a}})},\ \Eprint {https://arxiv.org/abs/1511.08687}
  {arXiv:1511.08687 [hep-th]} \BibitemShut {NoStop}%
\bibitem [{\citenamefont {Donnay}\ \emph
  {et~al.}(2016{\natexlab{b}})\citenamefont {Donnay}, \citenamefont {Giribet},
  \citenamefont {Gonz{\'a}lez},\ and\ \citenamefont {Pino}}]{Donnay:2016ejv}%
  \BibitemOpen
  \bibfield  {author} {\bibinfo {author} {\bibfnamefont {L.}~\bibnamefont
  {Donnay}}, \bibinfo {author} {\bibfnamefont {G.}~\bibnamefont {Giribet}},
  \bibinfo {author} {\bibfnamefont {H.~A.}\ \bibnamefont {Gonz{\'a}lez}},\ and\
  \bibinfo {author} {\bibfnamefont {M.}~\bibnamefont {Pino}},\ }\bibfield
  {title} {\bibinfo {title} {{Extended Symmetries at the Black Hole Horizon}},\
  }\href {https://doi.org/10.1007/JHEP09(2016)100} {\bibfield  {journal}
  {\bibinfo  {journal} {JHEP}\ }\textbf {\bibinfo {volume} {09}},\ \bibinfo
  {pages} {100} (2016)},\ \Eprint {https://arxiv.org/abs/1607.05703} {arXiv:1607.05703
  [hep-th]} \BibitemShut {NoStop}%
\bibitem [{\citenamefont {Ashtekar}\ and\ \citenamefont
  {Speziale}(2024{\natexlab{a}})}]{Ashtekar:2024mme}%
  \BibitemOpen
  \bibfield  {author} {\bibinfo {author} {\bibfnamefont {A.}~\bibnamefont
  {Ashtekar}}\ and\ \bibinfo {author} {\bibfnamefont {S.}~\bibnamefont
  {Speziale}},\ }\bibfield  {title} {\bibinfo {title} {{Horizons and null
  infinity: A fugue in four voices}},\ }\href
  {https://doi.org/10.1103/PhysRevD.109.L061501} {\bibfield  {journal}
  {\bibinfo  {journal} {Phys. Rev. D}\ }\textbf {\bibinfo {volume} {109}},\
  \bibinfo {pages} {L061501} (\bibinfo {year} {2024}{\natexlab{a}})},\ \Eprint
  {https://arxiv.org/abs/2401.15618} {arXiv:2401.15618 [gr-qc]} \BibitemShut
  {NoStop}%
\bibitem [{\citenamefont {Ashtekar}\ and\ \citenamefont
  {Speziale}(2024{\natexlab{b}})}]{Ashtekar:2024bpi}%
  \BibitemOpen
  \bibfield  {author} {\bibinfo {author} {\bibfnamefont {A.}~\bibnamefont
  {Ashtekar}}\ and\ \bibinfo {author} {\bibfnamefont {S.}~\bibnamefont
  {Speziale}},\ }\bibfield  {title} {\bibinfo {title} {{Null infinity as a
  weakly isolated horizon}},\ }\href
  {https://doi.org/10.1103/PhysRevD.110.044048} {\bibfield  {journal} {\bibinfo
   {journal} {Phys. Rev. D}\ }\textbf {\bibinfo {volume} {110}},\ \bibinfo
  {pages} {044048} (\bibinfo {year} {2024}{\natexlab{b}})},\ \Eprint
  {https://arxiv.org/abs/2402.17977} {arXiv:2402.17977 [hep-th]} \BibitemShut
  {NoStop}%
\bibitem [{\citenamefont {Ashtekar}\ and\ \citenamefont
  {Speziale}(2024{\natexlab{c}})}]{Ashtekar:2024stm}%
  \BibitemOpen
  \bibfield  {author} {\bibinfo {author} {\bibfnamefont {A.}~\bibnamefont
  {Ashtekar}}\ and\ \bibinfo {author} {\bibfnamefont {S.}~\bibnamefont
  {Speziale}},\ }\bibfield  {title} {\bibinfo {title} {{Null infinity and
  horizons: A new approach to fluxes and charges}},\ }\href
  {https://doi.org/10.1103/PhysRevD.110.044049} {\bibfield  {journal} {\bibinfo
   {journal} {Phys. Rev. D}\ }\textbf {\bibinfo {volume} {110}},\ \bibinfo
  {pages} {044049} (\bibinfo {year} {2024}{\natexlab{c}})},\ \Eprint
  {https://arxiv.org/abs/2407.03254} {arXiv:2407.03254 [hep-th]} \BibitemShut
  {NoStop}%
\bibitem [{early()}]{early}%
  \BibitemOpen
  \href@noop {} {}\bibinfo {note} {\blue{It is important to emphasize that the bulk supertranslations constructed in this work correspond to configurations defined throughout the entire spacetime. While, supertranslations associated with near-horizon symmetry can be defined on any finite-distance null hypersurface \cite{Adami:2020ugu,Chandrasekaran:2021hxc,Adami:2021nnf}. Although such hypersurfaces may be chosen arbitrarily within the spacetime, the resulting symmetry is intrinsically tied to the chosen hypersurface and do not extend naturally away from it. This difference reflects a general limitation of approaches that rely on series expansions around a selected null hypersurface. In those frameworks one can analyze the symmetry structure order by order in the small parameter expansion, for example through charge constructions \cite{Conde:2016csj,Conde:2016rom,Godazgar:2018vmm} or the symplectic structure in the covariant phase space formalism \cite{Ciambelli:2025mex}, which provides valuable information about the local structure near the hypersurface. However, they do not directly yield a symmetry transformation defined globally in the bulk due to convergence constraints of small parameter expansions \cite{Friedrich:1982}. }}\BibitemShut {Stop}%
\bibitem [{\citenamefont {Adami}\ \emph
  {et~al.}(2020{\natexlab{b}})\citenamefont {Adami}, \citenamefont
  {Sheikh-Jabbari}, \citenamefont {Taghiloo}, \citenamefont {Yavartanoo},\ and\
  \citenamefont {Zwikel}}]{Adami:2020ugu}%
  \BibitemOpen
  \bibfield  {author} {\bibinfo {author} {\bibfnamefont {H.}~\bibnamefont
  {Adami}}, \bibinfo {author} {\bibfnamefont {M.~M.}\ \bibnamefont
  {Sheikh-Jabbari}}, \bibinfo {author} {\bibfnamefont {V.}~\bibnamefont
  {Taghiloo}}, \bibinfo {author} {\bibfnamefont {H.}~\bibnamefont
  {Yavartanoo}},\ and\ \bibinfo {author} {\bibfnamefont {C.}~\bibnamefont
  {Zwikel}},\ }\bibfield  {title} {\bibinfo {title} {{Symmetries at null
  boundaries: two and three dimensional gravity cases}},\ }\href
  {https://doi.org/10.1007/JHEP10(2020)107} {\bibfield  {journal} {\bibinfo
  {journal} {JHEP}\ }\textbf {\bibinfo {volume} {10}},\ \bibinfo {pages}
  {107} (2020)},\ \Eprint {https://arxiv.org/abs/2007.12759} {arXiv:2007.12759
  [hep-th]} \BibitemShut {NoStop}%
\bibitem [{\citenamefont {Chandrasekaran}\ \emph {et~al.}(2022)\citenamefont
  {Chandrasekaran}, \citenamefont {Flanagan}, \citenamefont {Shehzad},\ and\
  \citenamefont {Speranza}}]{Chandrasekaran:2021hxc}%
  \BibitemOpen
  \bibfield  {author} {\bibinfo {author} {\bibfnamefont {V.}~\bibnamefont
  {Chandrasekaran}}, \bibinfo {author} {\bibfnamefont {E.~E.}\ \bibnamefont
  {Flanagan}}, \bibinfo {author} {\bibfnamefont {I.}~\bibnamefont {Shehzad}},\
  and\ \bibinfo {author} {\bibfnamefont {A.~J.}\ \bibnamefont {Speranza}},\
  }\bibfield  {title} {\bibinfo {title} {{Brown-York charges at null
  boundaries}},\ }\href {https://doi.org/10.1007/JHEP01(2022)029} {\bibfield
  {journal} {\bibinfo  {journal} {JHEP}\ }\textbf {\bibinfo {volume} {01}},\
  \bibinfo {pages} {029} (2022)},\ \Eprint {https://arxiv.org/abs/2109.11567}
  {arXiv:2109.11567 [hep-th]} \BibitemShut {NoStop}%
\bibitem [{\citenamefont {Adami}\ \emph {et~al.}(2021)\citenamefont {Adami},
  \citenamefont {Grumiller}, \citenamefont {Sheikh-Jabbari}, \citenamefont
  {Taghiloo}, \citenamefont {Yavartanoo},\ and\ \citenamefont
  {Zwikel}}]{Adami:2021nnf}%
  \BibitemOpen
  \bibfield  {author} {\bibinfo {author} {\bibfnamefont {H.}~\bibnamefont
  {Adami}}, \bibinfo {author} {\bibfnamefont {D.}~\bibnamefont {Grumiller}},
  \bibinfo {author} {\bibfnamefont {M.~M.}\ \bibnamefont {Sheikh-Jabbari}},
  \bibinfo {author} {\bibfnamefont {V.}~\bibnamefont {Taghiloo}}, \bibinfo
  {author} {\bibfnamefont {H.}~\bibnamefont {Yavartanoo}},\ and\ \bibinfo
  {author} {\bibfnamefont {C.}~\bibnamefont {Zwikel}},\ }\bibfield  {title}
  {\bibinfo {title} {{Null boundary phase space: slicings, news {\&} memory}},\
  }\href {https://doi.org/10.1007/JHEP11(2021)155} {\bibfield  {journal}
  {\bibinfo  {journal} {JHEP}\ }\textbf {\bibinfo {volume} {11}},\ \bibinfo
  {pages} {155} (2021)},\ \Eprint {https://arxiv.org/abs/2110.04218} {arXiv:2110.04218
  [hep-th]} \BibitemShut {NoStop}%
\bibitem [{\citenamefont {Conde}\ and\ \citenamefont
  {Mao}(2017{\natexlab{a}})}]{Conde:2016csj}%
  \BibitemOpen
  \bibfield  {author} {\bibinfo {author} {\bibfnamefont {E.}~\bibnamefont
  {Conde}}\ and\ \bibinfo {author} {\bibfnamefont {P.}~\bibnamefont {Mao}},\
  }\bibfield  {title} {\bibinfo {title} {{Remarks on asymptotic symmetries and
  the subleading soft photon theorem}},\ }\href
  {https://doi.org/10.1103/PhysRevD.95.021701} {\bibfield  {journal} {\bibinfo
  {journal} {Phys. Rev. D}\ }\textbf {\bibinfo {volume} {95}},\ \bibinfo
  {pages} {021701} (\bibinfo {year} {2017}{\natexlab{a}})},\ \Eprint
  {https://arxiv.org/abs/1605.09731} {arXiv:1605.09731 [hep-th]} \BibitemShut
  {NoStop}%
\bibitem [{\citenamefont {Conde}\ and\ \citenamefont
  {Mao}(2017{\natexlab{b}})}]{Conde:2016rom}%
  \BibitemOpen
  \bibfield  {author} {\bibinfo {author} {\bibfnamefont {E.}~\bibnamefont
  {Conde}}\ and\ \bibinfo {author} {\bibfnamefont {P.}~\bibnamefont {Mao}},\
  }\bibfield  {title} {\bibinfo {title} {{BMS Supertranslations and Not So Soft
  Gravitons}},\ }\href {https://doi.org/10.1007/JHEP05(2017)060} {\bibfield
  {journal} {\bibinfo  {journal} {JHEP}\ }\textbf {\bibinfo {volume} {05}},\
  \bibinfo {pages} {060} (2017)},\ \Eprint {https://arxiv.org/abs/1612.08294}
  {arXiv:1612.08294 [hep-th]} \BibitemShut {NoStop}%
\bibitem [{\citenamefont {Godazgar}\ \emph {et~al.}(2019)\citenamefont
  {Godazgar}, \citenamefont {Godazgar},\ and\ \citenamefont
  {Pope}}]{Godazgar:2018vmm}%
  \BibitemOpen
  \bibfield  {author} {\bibinfo {author} {\bibfnamefont {H.}~\bibnamefont
  {Godazgar}}, \bibinfo {author} {\bibfnamefont {M.}~\bibnamefont {Godazgar}},\
  and\ \bibinfo {author} {\bibfnamefont {C.~N.}\ \bibnamefont {Pope}},\
  }\bibfield  {title} {\bibinfo {title} {{Subleading BMS charges and fake news
  near null infinity}},\ }\href {https://doi.org/10.1007/JHEP01(2019)143}
  {\bibfield  {journal} {\bibinfo  {journal} {JHEP}\ }\textbf {\bibinfo
  {volume} {01}},\ \bibinfo {pages} {143} (2019)},\ \Eprint
  {https://arxiv.org/abs/1809.09076} {arXiv:1809.09076 [hep-th]} \BibitemShut
  {NoStop}%
\bibitem [{\citenamefont {Ciambelli}(2025)}]{Ciambelli:2025mex}%
  \BibitemOpen
  \bibfield  {author} {\bibinfo {author} {\bibfnamefont {L.}~\bibnamefont
  {Ciambelli}},\ }\bibfield  {title} {\bibinfo {title} {{Asymptotic limit of
  null hypersurfaces}},\ }\href {https://doi.org/10.1088/1361-6382/ae22b5}
  {\bibfield  {journal} {\bibinfo  {journal} {Class. Quant. Grav.}\ }\textbf
  {\bibinfo {volume} {42}},\ \bibinfo {pages} {235020} (\bibinfo {year}
  {2025})},\ \Eprint {https://arxiv.org/abs/2501.17357} {arXiv:2501.17357
  [hep-th]} \BibitemShut {NoStop}%
\bibitem [{\citenamefont {Friedrich}(1982)}]{Friedrich:1982}%
  \BibitemOpen
  \bibfield  {author} {\bibinfo {author} {\bibfnamefont {H.}~\bibnamefont
  {Friedrich}},\ }\bibfield  {title} {\bibinfo {title} {{On the Existence of
  Analytic Null Asymptotically Flat Solutions of Einstein’s Vacuum Field
  Equations}},\ }\href {https://doi.org/10.1098/rspa.1982.0077} {\bibfield
  {journal} {\bibinfo  {journal} {Proc. Roy. Soc. Lond.}\ }\textbf {\bibinfo
  {volume} {A381}},\ \bibinfo {pages} {361} (\bibinfo {year}
  {1982})}\BibitemShut {NoStop}%
\bibitem [{\citenamefont {Hawking}(2015)}]{Hawking:2015qqa}%
  \BibitemOpen
  \bibfield  {author} {\bibinfo {author} {\bibfnamefont {S.~W.}\ \bibnamefont
  {Hawking}},\ }\bibfield  {title} {\bibinfo {title} {{The Information Paradox
  for Black Holes}}\ }(\bibinfo {year} {2015})\ \Eprint
  {https://arxiv.org/abs/1509.01147} {arXiv:1509.01147 [hep-th]} \BibitemShut
  {NoStop}%
\bibitem [{\citenamefont {Hawking}\ \emph {et~al.}(2016)\citenamefont
  {Hawking}, \citenamefont {Perry},\ and\ \citenamefont
  {Strominger}}]{Hawking:2016msc}%
  \BibitemOpen
  \bibfield  {author} {\bibinfo {author} {\bibfnamefont {S.~W.}\ \bibnamefont
  {Hawking}}, \bibinfo {author} {\bibfnamefont {M.~J.}\ \bibnamefont {Perry}},\
  and\ \bibinfo {author} {\bibfnamefont {A.}~\bibnamefont {Strominger}},\
  }\bibfield  {title} {\bibinfo {title} {{Soft Hair on Black Holes}},\ }\href
  {https://doi.org/10.1103/PhysRevLett.116.231301} {\bibfield  {journal}
  {\bibinfo  {journal} {Phys. Rev. Lett.}\ }\textbf {\bibinfo {volume} {116}},\
  \bibinfo {pages} {231301} (\bibinfo {year} {2016})},\ \Eprint
  {https://arxiv.org/abs/1601.00921} {arXiv:1601.00921 [hep-th]} \BibitemShut
  {NoStop}%
\bibitem [{\citenamefont {Bart}(2020)}]{Bart:2019gnf}%
  \BibitemOpen
  \bibfield  {author} {\bibinfo {author} {\bibfnamefont {H.}~\bibnamefont
  {Bart}},\ }\bibfield  {title} {\bibinfo {title} {{Gravitational memory in the
  bulk}},\ }\href {https://doi.org/10.1007/JHEP05(2020)106} {\bibfield
  {journal} {\bibinfo  {journal} {JHEP}\ }\textbf {\bibinfo {volume} {05}},\
  \bibinfo {pages} {106} (2020)},\ \Eprint {https://arxiv.org/abs/1908.07505}
  {arXiv:1908.07505 [gr-qc]} \BibitemShut {NoStop}%
\bibitem [{\citenamefont {Newman}\ and\ \citenamefont
  {Penrose}(1962)}]{Newman:1961qr}%
  \BibitemOpen
  \bibfield  {author} {\bibinfo {author} {\bibfnamefont {E.}~\bibnamefont
  {Newman}}\ and\ \bibinfo {author} {\bibfnamefont {R.}~\bibnamefont
  {Penrose}},\ }\bibfield  {title} {\bibinfo {title} {{An Approach to
  gravitational radiation by a method of spin coefficients}},\ }\href
  {https://doi.org/10.1063/1.1724257} {\bibfield  {journal} {\bibinfo
  {journal} {J. Math. Phys.}\ }\textbf {\bibinfo {volume} {3}},\ \bibinfo
  {pages} {566} (\bibinfo {year} {1962})}\BibitemShut {NoStop}%
\bibitem [{\citenamefont {Newman}\ and\ \citenamefont
  {Unti}(1962)}]{Newman:1962cia}%
  \BibitemOpen
  \bibfield  {author} {\bibinfo {author} {\bibfnamefont {E.~T.}\ \bibnamefont
  {Newman}}\ and\ \bibinfo {author} {\bibfnamefont {T.~W.~J.}\ \bibnamefont
  {Unti}},\ }\bibfield  {title} {\bibinfo {title} {{Behavior of Asymptotically
  Flat Empty Spaces}},\ }\href {https://doi.org/10.1063/1.1724303} {\bibfield
  {journal} {\bibinfo  {journal} {J. Math. Phys.}\ }\textbf {\bibinfo {volume}
  {3}},\ \bibinfo {pages} {891} (\bibinfo {year} {1962})}\BibitemShut {NoStop}%
\bibitem [{\citenamefont {Barnich}\ and\ \citenamefont
  {Lambert}(2012)}]{Barnich:2011ty}%
  \BibitemOpen
  \bibfield  {author} {\bibinfo {author} {\bibfnamefont {G.}~\bibnamefont
  {Barnich}}\ and\ \bibinfo {author} {\bibfnamefont {P.-H.}\ \bibnamefont
  {Lambert}},\ }\bibfield  {title} {\bibinfo {title} {{A Note on the
  Newman-Unti group and the BMS charge algebra in terms of Newman-Penrose
  coefficients}},\ }\href {https://doi.org/10.1155/2012/197385} {\bibfield
  {journal} {\bibinfo  {journal} {Adv. Math. Phys.}\ }\textbf {\bibinfo
  {volume} {2012}},\ \bibinfo {pages} {197385} (\bibinfo {year} {2012})},\
  \Eprint {https://arxiv.org/abs/1102.0589} {arXiv:1102.0589 [gr-qc]}
  \BibitemShut {NoStop}%
\bibitem [{\citenamefont {Geiller}\ and\ \citenamefont
  {Zwikel}(2022)}]{Geiller:2022vto}%
  \BibitemOpen
  \bibfield  {author} {\bibinfo {author} {\bibfnamefont {M.}~\bibnamefont
  {Geiller}}\ and\ \bibinfo {author} {\bibfnamefont {C.}~\bibnamefont
  {Zwikel}},\ }\bibfield  {title} {\bibinfo {title} {{The partial Bondi gauge:
  Further enlarging the asymptotic structure of gravity}},\ }\href
  {https://doi.org/10.21468/SciPostPhys.13.5.108} {\bibfield  {journal}
  {\bibinfo  {journal} {SciPost Phys.}\ }\textbf {\bibinfo {volume} {13}},\
  \bibinfo {pages} {108} (\bibinfo {year} {2022})},\ \Eprint
  {https://arxiv.org/abs/2205.11401} {arXiv:2205.11401 [hep-th]} \BibitemShut
  {NoStop}%
\bibitem [{\citenamefont {Geiller}\ and\ \citenamefont
  {Zwikel}(2024)}]{Geiller:2024amx}%
  \BibitemOpen
  \bibfield  {author} {\bibinfo {author} {\bibfnamefont {M.}~\bibnamefont
  {Geiller}}\ and\ \bibinfo {author} {\bibfnamefont {C.}~\bibnamefont
  {Zwikel}},\ }\bibfield  {title} {\bibinfo {title} {{The partial Bondi gauge:
  Gauge fixings and asymptotic charges}},\ }\href
  {https://doi.org/10.21468/SciPostPhys.16.3.076} {\bibfield  {journal}
  {\bibinfo  {journal} {SciPost Phys.}\ }\textbf {\bibinfo {volume} {16}},\
  \bibinfo {pages} {076} (\bibinfo {year} {2024})},\ \Eprint
  {https://arxiv.org/abs/2401.09540} {arXiv:2401.09540 [hep-th]} \BibitemShut
  {NoStop}%
\bibitem [{\citenamefont {Penrose}(1963)}]{Penrose:1962ij}%
  \BibitemOpen
  \bibfield  {author} {\bibinfo {author} {\bibfnamefont {R.}~\bibnamefont
  {Penrose}},\ }\bibfield  {title} {\bibinfo {title} {{Asymptotic properties of
  fields and space-times}},\ }\href {https://doi.org/10.1103/PhysRevLett.10.66}
  {\bibfield  {journal} {\bibinfo  {journal} {Phys. Rev. Lett.}\ }\textbf
  {\bibinfo {volume} {10}},\ \bibinfo {pages} {66} (\bibinfo {year}
  {1963})}\BibitemShut {NoStop}%
\bibitem [{\citenamefont {Geroch}(1977)}]{Geroch:1977big}%
  \BibitemOpen
  \bibfield  {author} {\bibinfo {author} {\bibfnamefont {R.}~\bibnamefont
  {Geroch}},\ }\bibfield  {title} {\bibinfo {title} {{Asymptotic Structure of
  Space-Time}},\ }in\ \href {https://doi.org/10.1007/978-1-4684-2343-3_1}
  {\emph {\bibinfo {booktitle} {{Symposium on Asymptotic Structure of
  Space-Time}}}}\ (\bibinfo {year} {1977})\BibitemShut {NoStop}%
\bibitem [{\citenamefont {Wald}(1984)}]{Wald:1984rg}%
  \BibitemOpen
  \bibfield  {author} {\bibinfo {author} {\bibfnamefont {R.~M.}\ \bibnamefont
  {Wald}},\ }\href {https://doi.org/10.7208/chicago/9780226870373.001.0001}
  {\emph {\bibinfo {title} {{General Relativity}}}}\ (\bibinfo  {publisher}
  {Chicago Univ. Pr.},\ \bibinfo {address} {Chicago, USA},\ \bibinfo {year}
  {1984})\BibitemShut {NoStop}%
\bibitem [{\citenamefont {Ashtekar}(2015)}]{Ashtekar:2014zsa}%
  \BibitemOpen
  \bibfield  {author} {\bibinfo {author} {\bibfnamefont {A.}~\bibnamefont
  {Ashtekar}},\ }\bibfield  {title} {\bibinfo {title} {{Geometry and physics of
  null infinity}},\ }\href {https://doi.org/10.4310/sdg.2015.v20.n1.a5}
  {\bibfield  {journal} {\bibinfo  {journal} {Surveys Diff. Geom.}\ }\textbf
  {\bibinfo {volume} {20}},\ \bibinfo {pages} {99} (\bibinfo {year} {2015})},\
  \Eprint {https://arxiv.org/abs/1409.1800} {arXiv:1409.1800 [gr-qc]}
  \BibitemShut {NoStop}%
\bibitem [{\citenamefont {Goncharov}\ \emph {et~al.}(2024)\citenamefont
  {Goncharov}, \citenamefont {Donnay},\ and\ \citenamefont
  {Harms}}]{Goncharov:2023woe}%
  \BibitemOpen
  \bibfield  {author} {\bibinfo {author} {\bibfnamefont {B.}~\bibnamefont
  {Goncharov}}, \bibinfo {author} {\bibfnamefont {L.}~\bibnamefont {Donnay}},\
  and\ \bibinfo {author} {\bibfnamefont {J.}~\bibnamefont {Harms}},\ }\bibfield
   {title} {\bibinfo {title} {{Inferring Fundamental Spacetime Symmetries with
  Gravitational-Wave Memory: From LISA to the Einstein Telescope}},\ }\href
  {https://doi.org/10.1103/PhysRevLett.132.241401} {\bibfield  {journal}
  {\bibinfo  {journal} {Phys. Rev. Lett.}\ }\textbf {\bibinfo {volume} {132}},\
  \bibinfo {pages} {241401} (\bibinfo {year} {2024})},\ \Eprint
  {https://arxiv.org/abs/2310.10718} {arXiv:2310.10718 [gr-qc]} \BibitemShut
  {NoStop}%
\bibitem [{flat()}]{flat}%
  \BibitemOpen
  \href@noop {} {}\bibinfo {note} {In flat spacetime, light rays remain straight, so bulk memory (supertranslations) can distort their trajectories but never generate turning points. \red{It is important to emphasize that the curvature-dependent memory effect does not create turning points for null geodesics. The existence of turning points is an intrinsic feature of the black hole geometry. Instead, gravitational-wave memory induces a transition from null geodesics without turning points to null geodesics with turning points.}}\BibitemShut {Stop}%
\bibitem [{\citenamefont {Donnay}\ \emph {et~al.}(2018)\citenamefont {Donnay},
  \citenamefont {Giribet}, \citenamefont {Gonz{\'a}lez},\ and\ \citenamefont
  {Puhm}}]{Donnay:2018ckb}%
  \BibitemOpen
  \bibfield  {author} {\bibinfo {author} {\bibfnamefont {L.}~\bibnamefont
  {Donnay}}, \bibinfo {author} {\bibfnamefont {G.}~\bibnamefont {Giribet}},
  \bibinfo {author} {\bibfnamefont {H.~A.}\ \bibnamefont {Gonz{\'a}lez}},\ and\
  \bibinfo {author} {\bibfnamefont {A.}~\bibnamefont {Puhm}},\ }\bibfield
  {title} {\bibinfo {title} {{Black hole memory effect}},\ }\href
  {https://doi.org/10.1103/PhysRevD.98.124016} {\bibfield  {journal} {\bibinfo
  {journal} {Phys. Rev. D}\ }\textbf {\bibinfo {volume} {98}},\ \bibinfo
  {pages} {124016} (\bibinfo {year} {2018})},\ \Eprint
  {https://arxiv.org/abs/1809.07266} {arXiv:1809.07266 [hep-th]} \BibitemShut
  {NoStop}%
\bibitem [{\citenamefont {Rahman}\ and\ \citenamefont
  {Wald}(2020)}]{Rahman:2019bmk}%
  \BibitemOpen
  \bibfield  {author} {\bibinfo {author} {\bibfnamefont {A.~A.}\ \bibnamefont
  {Rahman}}\ and\ \bibinfo {author} {\bibfnamefont {R.~M.}\ \bibnamefont
  {Wald}},\ }\bibfield  {title} {\bibinfo {title} {{Black Hole Memory}},\
  }\href {https://doi.org/10.1103/PhysRevD.101.124010} {\bibfield  {journal}
  {\bibinfo  {journal} {Phys. Rev. D}\ }\textbf {\bibinfo {volume} {101}},\
  \bibinfo {pages} {124010} (\bibinfo {year} {2020})},\ \Eprint
  {https://arxiv.org/abs/1912.12806} {arXiv:1912.12806 [gr-qc]} \BibitemShut
  {NoStop}%
\bibitem [{\citenamefont {Elhashash}\ and\ \citenamefont
  {Nichols}(2025)}]{Elhashash:2024thm}%
  \BibitemOpen
  \bibfield  {author} {\bibinfo {author} {\bibfnamefont {A.}~\bibnamefont
  {Elhashash}}\ and\ \bibinfo {author} {\bibfnamefont {D.~A.}\ \bibnamefont
  {Nichols}},\ }\bibfield  {title} {\bibinfo {title} {{Waveform models for the
  gravitational-wave memory effect: Extreme mass-ratio limit and final memory
  offset}},\ }\href {https://doi.org/10.1103/PhysRevD.111.044052} {\bibfield
  {journal} {\bibinfo  {journal} {Phys. Rev. D}\ }\textbf {\bibinfo {volume}
  {111}},\ \bibinfo {pages} {044052} (\bibinfo {year} {2025})},\ \Eprint
  {https://arxiv.org/abs/2407.19017} {arXiv:2407.19017 [gr-qc]} \BibitemShut
  {NoStop}%
\bibitem [{\citenamefont {Cunningham}\ \emph {et~al.}(2025)\citenamefont
  {Cunningham}, \citenamefont {Kavanagh}, \citenamefont {Pound}, \citenamefont
  {Trestini}, \citenamefont {Warburton},\ and\ \citenamefont
  {Neef}}]{Cunningham:2024dog}%
  \BibitemOpen
  \bibfield  {author} {\bibinfo {author} {\bibfnamefont {K.}~\bibnamefont
  {Cunningham}}, \bibinfo {author} {\bibfnamefont {C.}~\bibnamefont
  {Kavanagh}}, \bibinfo {author} {\bibfnamefont {A.}~\bibnamefont {Pound}},
  \bibinfo {author} {\bibfnamefont {D.}~\bibnamefont {Trestini}}, \bibinfo
  {author} {\bibfnamefont {N.}~\bibnamefont {Warburton}},\ and\ \bibinfo
  {author} {\bibfnamefont {J.}~\bibnamefont {Neef}},\ }\bibfield  {title}
  {\bibinfo {title} {{Gravitational memory: new results from post-Newtonian and
  self-force theory}},\ }\href {https://doi.org/10.1088/1361-6382/adbc3d}
  {\bibfield  {journal} {\bibinfo  {journal} {Class. Quant. Grav.}\ }\textbf
  {\bibinfo {volume} {42}},\ \bibinfo {pages} {135009} (\bibinfo {year}
  {2025})},\ \bibinfo {note} {[Addendum: Class.Quant.Grav. 42, 199401
  (2025)]},\ \Eprint {https://arxiv.org/abs/2410.23950} {arXiv:2410.23950
  [gr-qc]} \BibitemShut {NoStop}%
\bibitem [{\citenamefont {Brodsky}\ \emph {et~al.}(1998)\citenamefont
  {Brodsky}, \citenamefont {Pauli},\ and\ \citenamefont
  {Pinsky}}]{Brodsky:1997de}%
  \BibitemOpen
  \bibfield  {author} {\bibinfo {author} {\bibfnamefont {S.~J.}\ \bibnamefont
  {Brodsky}}, \bibinfo {author} {\bibfnamefont {H.-C.}\ \bibnamefont {Pauli}},\
  and\ \bibinfo {author} {\bibfnamefont {S.~S.}\ \bibnamefont {Pinsky}},\
  }\bibfield  {title} {\bibinfo {title} {{Quantum chromodynamics and other
  field theories on the light cone}},\ }\href
  {https://doi.org/10.1016/S0370-1573(97)00089-6} {\bibfield  {journal}
  {\bibinfo  {journal} {Phys. Rept.}\ }\textbf {\bibinfo {volume} {301}},\
  \bibinfo {pages} {299} (\bibinfo {year} {1998})},\ \Eprint
  {https://arxiv.org/abs/hep-ph/9705477} {arXiv:hep-ph/9705477} \BibitemShut
  {NoStop}%
\bibitem [{\citenamefont {Heinzl}(2001)}]{Heinzl:2000ht}%
  \BibitemOpen
  \bibfield  {author} {\bibinfo {author} {\bibfnamefont {T.}~\bibnamefont
  {Heinzl}},\ }\bibfield  {title} {\bibinfo {title} {{Light cone quantization:
  Foundations and applications}},\ }\href
  {https://doi.org/10.1007/3-540-45114-5_2} {\bibfield  {journal} {\bibinfo
  {journal} {Lect. Notes Phys.}\ }\textbf {\bibinfo {volume} {572}},\ \bibinfo
  {pages} {55} (\bibinfo {year} {2001})},\ \Eprint
  {https://arxiv.org/abs/hep-th/0008096} {arXiv:hep-th/0008096} \BibitemShut
  {NoStop}%
\bibitem [{hypersurface()}]{hypersurface}%
  \BibitemOpen
  \href@noop {} {}\bibinfo {note} {There are freedoms on the null hypersurface from re-parameterizing null geodesics and choosing the remaining two coordinates $x^a$. \blue{Nevertheless, those freedoms are kinematics associated to chosen null hypersurface.} In the present work, we focus on transformations that induce physical transitions between null hypersurfaces.}\BibitemShut {Stop}%
\bibitem [{\citenamefont {Winicour}(2001)}]{Winicour:2001imp}%
  \BibitemOpen
  \bibfield  {author} {\bibinfo {author} {\bibfnamefont {J.}~\bibnamefont
  {Winicour}},\ }\bibfield  {title} {\bibinfo {title} {{Characteristic
  evolution and matching}},\ }\href {https://doi.org/10.12942/lrr-2001-3}
  {\bibfield  {journal} {\bibinfo  {journal} {Living Rev. Rel.}\ }\textbf
  {\bibinfo {volume} {4}},\ \bibinfo {pages} {3} (\bibinfo {year} {2001})},\
  \Eprint {https://arxiv.org/abs/gr-qc/0102085} {arXiv:gr-qc/0102085}
  \BibitemShut {NoStop}%
\bibitem [{\citenamefont {Chesler}\ and\ \citenamefont
  {Yaffe}(2014)}]{Chesler:2013lia}%
  \BibitemOpen
  \bibfield  {author} {\bibinfo {author} {\bibfnamefont {P.~M.}\ \bibnamefont
  {Chesler}}\ and\ \bibinfo {author} {\bibfnamefont {L.~G.}\ \bibnamefont
  {Yaffe}},\ }\bibfield  {title} {\bibinfo {title} {{Numerical solution of
  gravitational dynamics in asymptotically anti-de Sitter spacetimes}},\ }\href
  {https://doi.org/10.1007/JHEP07(2014)086} {\bibfield  {journal} {\bibinfo
  {journal} {JHEP}\ }\textbf {\bibinfo {volume} {07}},\ \bibinfo {pages}
  {086} (2014)},\ \Eprint {https://arxiv.org/abs/1309.1439} {arXiv:1309.1439 [hep-th]}
  \BibitemShut {NoStop}%
\bibitem [{\citenamefont {Barnich}\ and\ \citenamefont
  {Troessaert}(2016)}]{Barnich:2016lyg}%
  \BibitemOpen
  \bibfield  {author} {\bibinfo {author} {\bibfnamefont {G.}~\bibnamefont
  {Barnich}}\ and\ \bibinfo {author} {\bibfnamefont {C.}~\bibnamefont
  {Troessaert}},\ }\bibfield  {title} {\bibinfo {title} {{Finite BMS
  transformations}},\ }\href {https://doi.org/10.1007/JHEP03(2016)167}
  {\bibfield  {journal} {\bibinfo  {journal} {JHEP}\ }\textbf {\bibinfo
  {volume} {03}},\ \bibinfo {pages} {167} (2016)},\ \Eprint
  {https://arxiv.org/abs/1601.04090} {arXiv:1601.04090 [gr-qc]} \BibitemShut
  {NoStop}%
\bibitem [{\citenamefont {Flanagan}\ and\ \citenamefont
  {Nichols}(2024)}]{Flanagan:2023jio}%
  \BibitemOpen
  \bibfield  {author} {\bibinfo {author} {\bibfnamefont {E.~E.}\ \bibnamefont
  {Flanagan}}\ and\ \bibinfo {author} {\bibfnamefont {D.~A.}\ \bibnamefont
  {Nichols}},\ }\bibfield  {title} {\bibinfo {title} {{Fully nonlinear
  transformations of the Weyl-Bondi-Metzner-Sachs asymptotic symmetry group}},\
  }\href {https://doi.org/10.1007/JHEP03(2024)120} {\bibfield  {journal}
  {\bibinfo  {journal} {JHEP}\ }\textbf {\bibinfo {volume} {03}},\ \bibinfo
  {pages} {120} (2024)},\ \Eprint {https://arxiv.org/abs/2311.03130} {arXiv:2311.03130
  [gr-qc]} \BibitemShut {NoStop}%
\bibitem [{\citenamefont {Barnich}\ and\ \citenamefont
  {Troessaert}(2010{\natexlab{a}})}]{Barnich:2009se}%
  \BibitemOpen
  \bibfield  {author} {\bibinfo {author} {\bibfnamefont {G.}~\bibnamefont
  {Barnich}}\ and\ \bibinfo {author} {\bibfnamefont {C.}~\bibnamefont
  {Troessaert}},\ }\bibfield  {title} {\bibinfo {title} {{Symmetries of
  asymptotically flat 4 dimensional spacetimes at null infinity revisited}},\
  }\href {https://doi.org/10.1103/PhysRevLett.105.111103} {\bibfield  {journal}
  {\bibinfo  {journal} {Phys. Rev. Lett.}\ }\textbf {\bibinfo {volume} {105}},\
  \bibinfo {pages} {111103} (\bibinfo {year} {2010}{\natexlab{a}})},\ \Eprint
  {https://arxiv.org/abs/0909.2617} {arXiv:0909.2617 [gr-qc]} \BibitemShut
  {NoStop}%
\bibitem [{\citenamefont {Barnich}\ and\ \citenamefont
  {Troessaert}(2010{\natexlab{b}})}]{Barnich:2010eb}%
  \BibitemOpen
  \bibfield  {author} {\bibinfo {author} {\bibfnamefont {G.}~\bibnamefont
  {Barnich}}\ and\ \bibinfo {author} {\bibfnamefont {C.}~\bibnamefont
  {Troessaert}},\ }\bibfield  {title} {\bibinfo {title} {{Aspects of the
  BMS/CFT correspondence}},\ }\href {https://doi.org/10.1007/JHEP05(2010)062}
  {\bibfield  {journal} {\bibinfo  {journal} {JHEP}\ }\textbf {\bibinfo
  {volume} {05}},\ \bibinfo {pages} {062} (2010)},\ \Eprint
  {https://arxiv.org/abs/1001.1541} {arXiv:1001.1541 [hep-th]} \BibitemShut
  {NoStop}%
\bibitem [{\citenamefont {Barnich}\ and\ \citenamefont
  {Troessaert}(2010{\natexlab{c}})}]{Barnich:2010ojg}%
  \BibitemOpen
  \bibfield  {author} {\bibinfo {author} {\bibfnamefont {G.}~\bibnamefont
  {Barnich}}\ and\ \bibinfo {author} {\bibfnamefont {C.}~\bibnamefont
  {Troessaert}},\ }\bibfield  {title} {\bibinfo {title} {{Supertranslations
  call for superrotations}},\ }\href {https://doi.org/10.22323/1.127.0010}
  {\bibfield  {journal} {\bibinfo  {journal} {PoS}\ }\textbf {\bibinfo {volume}
  {CNCFG2010}},\ \bibinfo {pages} {010} (\bibinfo {year}
  {2010}{\natexlab{c}})},\ \Eprint {https://arxiv.org/abs/1102.4632}
  {arXiv:1102.4632 [gr-qc]} \BibitemShut {NoStop}%
\bibitem [{\citenamefont {Barnich}\ and\ \citenamefont
  {Troessaert}(2011)}]{Barnich:2011mi}%
  \BibitemOpen
  \bibfield  {author} {\bibinfo {author} {\bibfnamefont {G.}~\bibnamefont
  {Barnich}}\ and\ \bibinfo {author} {\bibfnamefont {C.}~\bibnamefont
  {Troessaert}},\ }\bibfield  {title} {\bibinfo {title} {{BMS charge
  algebra}},\ }\href {https://doi.org/10.1007/JHEP12(2011)105} {\bibfield
  {journal} {\bibinfo  {journal} {JHEP}\ }\textbf {\bibinfo {volume} {12}},\
  \bibinfo {pages} {105} (2011)},\ \Eprint {https://arxiv.org/abs/1106.0213}
  {arXiv:1106.0213 [hep-th]} \BibitemShut {NoStop}%
\bibitem [{\citenamefont {Comp{\`e}re}\ \emph {et~al.}(2018)\citenamefont
  {Comp{\`e}re}, \citenamefont {Fiorucci},\ and\ \citenamefont
  {Ruzziconi}}]{Compere:2018ylh}%
  \BibitemOpen
  \bibfield  {author} {\bibinfo {author} {\bibfnamefont {G.}~\bibnamefont
  {Comp{\`e}re}}, \bibinfo {author} {\bibfnamefont {A.}~\bibnamefont
  {Fiorucci}},\ and\ \bibinfo {author} {\bibfnamefont {R.}~\bibnamefont
  {Ruzziconi}},\ }\bibfield  {title} {\bibinfo {title} {{Superboost
  transitions, refraction memory and super-Lorentz charge algebra}},\ }\href
  {https://doi.org/10.1007/JHEP11(2018)200} {\bibfield  {journal} {\bibinfo
  {journal} {JHEP}\ }\textbf {\bibinfo {volume} {11}},\ \bibinfo {pages}
  {200} (2018)},\ \bibinfo {note} {[Erratum: JHEP 04, 172 (2020)]},\ \Eprint
  {https://arxiv.org/abs/1810.00377} {arXiv:1810.00377 [hep-th]} \BibitemShut
  {NoStop}%
\bibitem [{\citenamefont {Campiglia}\ and\ \citenamefont
  {Laddha}(2015)}]{Campiglia:2015yka}%
  \BibitemOpen
  \bibfield  {author} {\bibinfo {author} {\bibfnamefont {M.}~\bibnamefont
  {Campiglia}}\ and\ \bibinfo {author} {\bibfnamefont {A.}~\bibnamefont
  {Laddha}},\ }\bibfield  {title} {\bibinfo {title} {{New symmetries for the
  Gravitational S-matrix}},\ }\href {https://doi.org/10.1007/JHEP04(2015)076}
  {\bibfield  {journal} {\bibinfo  {journal} {JHEP}\ }\textbf {\bibinfo
  {volume} {04}},\ \bibinfo {pages} {076} (2015)},\ \Eprint
  {https://arxiv.org/abs/1502.02318} {arXiv:1502.02318 [hep-th]} \BibitemShut
  {NoStop}%
\bibitem [{\citenamefont {Campiglia}\ and\ \citenamefont
  {Peraza}(2020)}]{Campiglia:2020qvc}%
  \BibitemOpen
  \bibfield  {author} {\bibinfo {author} {\bibfnamefont {M.}~\bibnamefont
  {Campiglia}}\ and\ \bibinfo {author} {\bibfnamefont {J.}~\bibnamefont
  {Peraza}},\ }\bibfield  {title} {\bibinfo {title} {{Generalized BMS charge
  algebra}},\ }\href {https://doi.org/10.1103/PhysRevD.101.104039} {\bibfield
  {journal} {\bibinfo  {journal} {Phys. Rev. D}\ }\textbf {\bibinfo {volume}
  {101}},\ \bibinfo {pages} {104039} (\bibinfo {year} {2020})},\ \Eprint
  {https://arxiv.org/abs/2002.06691} {arXiv:2002.06691 [gr-qc]} \BibitemShut
  {NoStop}%
\bibitem [{\citenamefont {Flanagan}\ and\ \citenamefont
  {Nichols}(2017)}]{Flanagan:2015pxa}%
  \BibitemOpen
  \bibfield  {author} {\bibinfo {author} {\bibfnamefont {{\'E}.~{\'E}.}\
  \bibnamefont {Flanagan}}\ and\ \bibinfo {author} {\bibfnamefont {D.~A.}\
  \bibnamefont {Nichols}},\ }\bibfield  {title} {\bibinfo {title} {{Conserved
  charges of the extended Bondi-Metzner-Sachs algebra}},\ }\href
  {https://doi.org/10.1103/PhysRevD.95.044002} {\bibfield  {journal} {\bibinfo
  {journal} {Phys. Rev. D}\ }\textbf {\bibinfo {volume} {95}},\ \bibinfo
  {pages} {044002} (\bibinfo {year} {2017})},\ \bibinfo {note} {[Erratum:
  Phys.Rev.D 108, 069902 (2023)]},\ \Eprint {https://arxiv.org/abs/1510.03386}
  {arXiv:1510.03386 [hep-th]} \BibitemShut {NoStop}%
\bibitem [{\citenamefont {Freidel}\ \emph {et~al.}(2021)\citenamefont
  {Freidel}, \citenamefont {Oliveri}, \citenamefont {Pranzetti},\ and\
  \citenamefont {Speziale}}]{Freidel:2021fxf}%
  \BibitemOpen
  \bibfield  {author} {\bibinfo {author} {\bibfnamefont {L.}~\bibnamefont
  {Freidel}}, \bibinfo {author} {\bibfnamefont {R.}~\bibnamefont {Oliveri}},
  \bibinfo {author} {\bibfnamefont {D.}~\bibnamefont {Pranzetti}},\ and\
  \bibinfo {author} {\bibfnamefont {S.}~\bibnamefont {Speziale}},\ }\bibfield
  {title} {\bibinfo {title} {{The Weyl BMS group and Einstein{\textquoteright}s
  equations}},\ }\href {https://doi.org/10.1007/JHEP07(2021)170} {\bibfield
  {journal} {\bibinfo  {journal} {JHEP}\ }\textbf {\bibinfo {volume} {07}},\
  \bibinfo {pages} {170} (2021)},\ \Eprint {https://arxiv.org/abs/2104.05793}
  {arXiv:2104.05793 [hep-th]} \BibitemShut {NoStop}%
\bibitem [{\citenamefont {Comp{\`e}re}\ \emph {et~al.}(2019)\citenamefont
  {Comp{\`e}re}, \citenamefont {Fiorucci},\ and\ \citenamefont
  {Ruzziconi}}]{Compere:2019bua}%
  \BibitemOpen
  \bibfield  {author} {\bibinfo {author} {\bibfnamefont {G.}~\bibnamefont
  {Comp{\`e}re}}, \bibinfo {author} {\bibfnamefont {A.}~\bibnamefont
  {Fiorucci}},\ and\ \bibinfo {author} {\bibfnamefont {R.}~\bibnamefont
  {Ruzziconi}},\ }\bibfield  {title} {\bibinfo {title} {{The $\Lambda$-BMS$_4$
  group of dS$_4$ and new boundary conditions for AdS$_4$}},\ }\href
  {https://doi.org/10.1088/1361-6382/ab3d4b} {\bibfield  {journal} {\bibinfo
  {journal} {Class. Quant. Grav.}\ }\textbf {\bibinfo {volume} {36}},\ \bibinfo
  {pages} {195017} (\bibinfo {year} {2019})},\ \bibinfo {note} {[Erratum:
  Class.Quant.Grav. 38, 229501 (2021)]},\ \Eprint
  {https://arxiv.org/abs/1905.00971} {arXiv:1905.00971 [gr-qc]} \BibitemShut
  {NoStop}%
\bibitem [{\citenamefont {Comp{\`e}re}\ \emph {et~al.}(2020)\citenamefont
  {Comp{\`e}re}, \citenamefont {Fiorucci},\ and\ \citenamefont
  {Ruzziconi}}]{Compere:2020lrt}%
  \BibitemOpen
  \bibfield  {author} {\bibinfo {author} {\bibfnamefont {G.}~\bibnamefont
  {Comp{\`e}re}}, \bibinfo {author} {\bibfnamefont {A.}~\bibnamefont
  {Fiorucci}},\ and\ \bibinfo {author} {\bibfnamefont {R.}~\bibnamefont
  {Ruzziconi}},\ }\bibfield  {title} {\bibinfo {title} {{The $\Lambda$-BMS$_4$
  charge algebra}},\ }\href {https://doi.org/10.1007/JHEP10(2020)205}
  {\bibfield  {journal} {\bibinfo  {journal} {JHEP}\ }\textbf {\bibinfo
  {volume} {10}},\ \bibinfo {pages} {205} (2020)},\ \Eprint
  {https://arxiv.org/abs/2004.10769} {arXiv:2004.10769 [hep-th]} \BibitemShut
  {NoStop}%
\bibitem [{\citenamefont {Afshar}\ \emph {et~al.}(2016)\citenamefont {Afshar},
  \citenamefont {Detournay}, \citenamefont {Grumiller}, \citenamefont {Merbis},
  \citenamefont {Perez}, \citenamefont {Tempo},\ and\ \citenamefont
  {Troncoso}}]{Afshar:2016wfy}%
  \BibitemOpen
  \bibfield  {author} {\bibinfo {author} {\bibfnamefont {H.}~\bibnamefont
  {Afshar}}, \bibinfo {author} {\bibfnamefont {S.}~\bibnamefont {Detournay}},
  \bibinfo {author} {\bibfnamefont {D.}~\bibnamefont {Grumiller}}, \bibinfo
  {author} {\bibfnamefont {W.}~\bibnamefont {Merbis}}, \bibinfo {author}
  {\bibfnamefont {A.}~\bibnamefont {Perez}}, \bibinfo {author} {\bibfnamefont
  {D.}~\bibnamefont {Tempo}},\ and\ \bibinfo {author} {\bibfnamefont
  {R.}~\bibnamefont {Troncoso}},\ }\bibfield  {title} {\bibinfo {title} {{Soft
  Heisenberg hair on black holes in three dimensions}},\ }\href
  {https://doi.org/10.1103/PhysRevD.93.101503} {\bibfield  {journal} {\bibinfo
  {journal} {Phys. Rev. D}\ }\textbf {\bibinfo {volume} {93}},\ \bibinfo
  {pages} {101503} (\bibinfo {year} {2016})},\ \Eprint
  {https://arxiv.org/abs/1603.04824} {arXiv:1603.04824 [hep-th]} \BibitemShut
  {NoStop}%
\bibitem [{\citenamefont {Shi}\ and\ \citenamefont {Mei}(2017)}]{Shi:2016jtn}%
  \BibitemOpen
  \bibfield  {author} {\bibinfo {author} {\bibfnamefont {C.}~\bibnamefont
  {Shi}}\ and\ \bibinfo {author} {\bibfnamefont {J.}~\bibnamefont {Mei}},\
  }\bibfield  {title} {\bibinfo {title} {{Extended Symmetries at Black Hole
  Horizons in Generic Dimensions}},\ }\href
  {https://doi.org/10.1103/PhysRevD.95.104053} {\bibfield  {journal} {\bibinfo
  {journal} {Phys. Rev. D}\ }\textbf {\bibinfo {volume} {95}},\ \bibinfo
  {pages} {104053} (\bibinfo {year} {2017})},\ \Eprint
  {https://arxiv.org/abs/1611.09491} {arXiv:1611.09491 [gr-qc]} \BibitemShut
  {NoStop}%
\bibitem [{\citenamefont {Akhmedov}\ and\ \citenamefont
  {Godazgar}(2017)}]{Akhmedov:2017ftb}%
  \BibitemOpen
  \bibfield  {author} {\bibinfo {author} {\bibfnamefont {E.~T.}\ \bibnamefont
  {Akhmedov}}\ and\ \bibinfo {author} {\bibfnamefont {M.}~\bibnamefont
  {Godazgar}},\ }\bibfield  {title} {\bibinfo {title} {{Symmetries at the black
  hole horizon}},\ }\href {https://doi.org/10.1103/PhysRevD.96.104025}
  {\bibfield  {journal} {\bibinfo  {journal} {Phys. Rev. D}\ }\textbf {\bibinfo
  {volume} {96}},\ \bibinfo {pages} {104025} (\bibinfo {year} {2017})},\
  \Eprint {https://arxiv.org/abs/1707.05517} {arXiv:1707.05517 [hep-th]}
  \BibitemShut {NoStop}%
\bibitem [{\citenamefont {Chandrasekaran}\ \emph {et~al.}(2018)\citenamefont
  {Chandrasekaran}, \citenamefont {Flanagan},\ and\ \citenamefont
  {Prabhu}}]{Chandrasekaran:2018aop}%
  \BibitemOpen
  \bibfield  {author} {\bibinfo {author} {\bibfnamefont {V.}~\bibnamefont
  {Chandrasekaran}}, \bibinfo {author} {\bibfnamefont {{\'E}.~{\'E}.}\
  \bibnamefont {Flanagan}},\ and\ \bibinfo {author} {\bibfnamefont
  {K.}~\bibnamefont {Prabhu}},\ }\bibfield  {title} {\bibinfo {title}
  {{Symmetries and charges of general relativity at null boundaries}},\ }\href
  {https://doi.org/10.1007/JHEP11(2018)125} {\bibfield  {journal} {\bibinfo
  {journal} {JHEP}\ }\textbf {\bibinfo {volume} {11}},\ \bibinfo {pages}
  {125} (2018)},\ \bibinfo {note} {[Erratum: JHEP 07, 224 (2023)]},\ \Eprint
  {https://arxiv.org/abs/1807.11499} {arXiv:1807.11499 [hep-th]} \BibitemShut
  {NoStop}%
\bibitem [{\citenamefont {Haco}\ \emph {et~al.}(2018)\citenamefont {Haco},
  \citenamefont {Hawking}, \citenamefont {Perry},\ and\ \citenamefont
  {Strominger}}]{Haco:2018ske}%
  \BibitemOpen
  \bibfield  {author} {\bibinfo {author} {\bibfnamefont {S.}~\bibnamefont
  {Haco}}, \bibinfo {author} {\bibfnamefont {S.~W.}\ \bibnamefont {Hawking}},
  \bibinfo {author} {\bibfnamefont {M.~J.}\ \bibnamefont {Perry}},\ and\
  \bibinfo {author} {\bibfnamefont {A.}~\bibnamefont {Strominger}},\ }\bibfield
   {title} {\bibinfo {title} {{Black Hole Entropy and Soft Hair}},\ }\href
  {https://doi.org/10.1007/JHEP12(2018)098} {\bibfield  {journal} {\bibinfo
  {journal} {JHEP}\ }\textbf {\bibinfo {volume} {12}},\ \bibinfo {pages}
  {098} (2018)},\ \Eprint {https://arxiv.org/abs/1810.01847} {arXiv:1810.01847
  [hep-th]} \BibitemShut {NoStop}%
\bibitem [{\citenamefont {Grumiller}\ \emph
  {et~al.}(2020{\natexlab{a}})\citenamefont {Grumiller}, \citenamefont
  {P{\'e}rez}, \citenamefont {Sheikh-Jabbari}, \citenamefont {Troncoso},\ and\
  \citenamefont {Zwikel}}]{Grumiller:2019fmp}%
  \BibitemOpen
  \bibfield  {author} {\bibinfo {author} {\bibfnamefont {D.}~\bibnamefont
  {Grumiller}}, \bibinfo {author} {\bibfnamefont {A.}~\bibnamefont
  {P{\'e}rez}}, \bibinfo {author} {\bibfnamefont {M.~M.}\ \bibnamefont
  {Sheikh-Jabbari}}, \bibinfo {author} {\bibfnamefont {R.}~\bibnamefont
  {Troncoso}},\ and\ \bibinfo {author} {\bibfnamefont {C.}~\bibnamefont
  {Zwikel}},\ }\bibfield  {title} {\bibinfo {title} {{Spacetime structure near
  generic horizons and soft hair}},\ }\href
  {https://doi.org/10.1103/PhysRevLett.124.041601} {\bibfield  {journal}
  {\bibinfo  {journal} {Phys. Rev. Lett.}\ }\textbf {\bibinfo {volume} {124}},\
  \bibinfo {pages} {041601} (\bibinfo {year} {2020}{\natexlab{a}})},\ \Eprint
  {https://arxiv.org/abs/1908.09833} {arXiv:1908.09833 [hep-th]} \BibitemShut
  {NoStop}%
\bibitem [{\citenamefont {Adami}\ \emph
  {et~al.}(2020{\natexlab{c}})\citenamefont {Adami}, \citenamefont
  {Hosseinzadeh},\ and\ \citenamefont {Sheikh-Jabbari}}]{Adami:2020uwd}%
  \BibitemOpen
  \bibfield  {author} {\bibinfo {author} {\bibfnamefont {H.}~\bibnamefont
  {Adami}}, \bibinfo {author} {\bibfnamefont {V.}~\bibnamefont
  {Hosseinzadeh}},\ and\ \bibinfo {author} {\bibfnamefont {M.~M.}\ \bibnamefont
  {Sheikh-Jabbari}},\ }\bibfield  {title} {\bibinfo {title} {{Sliding surface
  charges on AdS$_3$}},\ }\href
  {https://doi.org/10.1016/j.physletb.2020.135503} {\bibfield  {journal}
  {\bibinfo  {journal} {Phys. Lett. B}\ }\textbf {\bibinfo {volume} {806}},\
  \bibinfo {pages} {135503} (\bibinfo {year} {2020}{\natexlab{c}})},\ \Eprint
  {https://arxiv.org/abs/2002.09962} {arXiv:2002.09962 [hep-th]} \BibitemShut
  {NoStop}%
\bibitem [{\citenamefont {Adami}\ \emph
  {et~al.}(2020{\natexlab{a}})\citenamefont {Adami}, \citenamefont {Grumiller},
  \citenamefont {Sadeghian}, \citenamefont {Sheikh-Jabbari},\ and\
  \citenamefont {Zwikel}}]{Adami:2020amw}%
  \BibitemOpen
  \bibfield  {author} {\bibinfo {author} {\bibfnamefont {H.}~\bibnamefont
  {Adami}}, \bibinfo {author} {\bibfnamefont {D.}~\bibnamefont {Grumiller}},
  \bibinfo {author} {\bibfnamefont {S.}~\bibnamefont {Sadeghian}}, \bibinfo
  {author} {\bibfnamefont {M.~M.}\ \bibnamefont {Sheikh-Jabbari}},\ and\
  \bibinfo {author} {\bibfnamefont {C.}~\bibnamefont {Zwikel}},\ }\bibfield
  {title} {\bibinfo {title} {{T-Witts from the horizon}},\ }\href
  {https://doi.org/10.1007/JHEP04(2020)128} {\bibfield  {journal} {\bibinfo
  {journal} {JHEP}\ }\textbf {\bibinfo {volume} {04}},\ \bibinfo {pages}
  {128} (2020)},\ \Eprint {https://arxiv.org/abs/2002.08346} {arXiv:2002.08346
  [hep-th]} \BibitemShut {NoStop}%
\bibitem [{\citenamefont {Chen}\ \emph {et~al.}(2020)\citenamefont {Chen},
  \citenamefont {Chua}, \citenamefont {Liu}, \citenamefont {Speranza},\ and\
  \citenamefont {Torres}}]{Chen:2020nyh}%
  \BibitemOpen
  \bibfield  {author} {\bibinfo {author} {\bibfnamefont {L.-Q.}\ \bibnamefont
  {Chen}}, \bibinfo {author} {\bibfnamefont {W.~Z.}\ \bibnamefont {Chua}},
  \bibinfo {author} {\bibfnamefont {S.}~\bibnamefont {Liu}}, \bibinfo {author}
  {\bibfnamefont {A.~J.}\ \bibnamefont {Speranza}},\ and\ \bibinfo {author}
  {\bibfnamefont {B.~d. S.~L.}\ \bibnamefont {Torres}},\ }\bibfield  {title}
  {\bibinfo {title} {{Virasoro hair and entropy for axisymmetric Killing
  horizons}},\ }\href {https://doi.org/10.1103/PhysRevLett.125.241302}
  {\bibfield  {journal} {\bibinfo  {journal} {Phys. Rev. Lett.}\ }\textbf
  {\bibinfo {volume} {125}},\ \bibinfo {pages} {241302} (\bibinfo {year}
  {2020})},\ \Eprint {https://arxiv.org/abs/2006.02430} {arXiv:2006.02430
  [hep-th]} \BibitemShut {NoStop}%
\bibitem [{\citenamefont {Grumiller}\ \emph {et~al.}(2020)\citenamefont
  {Grumiller}, \citenamefont {Sheikh-Jabbari}, \citenamefont {Troessaert},\
  and\ \citenamefont {Wutte}}]{Grumiller:2019ygj}%
  \BibitemOpen
  \bibfield  {author} {\bibinfo {author} {\bibfnamefont {D.}~\bibnamefont
  {Grumiller}}, \bibinfo {author} {\bibfnamefont {M.~M.}\ \bibnamefont
  {Sheikh-Jabbari}}, \bibinfo {author} {\bibfnamefont {C.}~\bibnamefont
  {Troessaert}},\ and\ \bibinfo {author} {\bibfnamefont {R.}~\bibnamefont
  {Wutte}},\ }\bibfield  {title} {\bibinfo {title} {{Interpolating Between
  Asymptotic and Near Horizon Symmetries}},\ }\href
  {https://doi.org/10.1007/JHEP03(2020)035} {\bibfield  {journal} {\bibinfo
  {journal} {JHEP}\ }\textbf {\bibinfo {volume} {03}},\ \bibinfo {pages}
  {035} (2020)},\ \Eprint {https://arxiv.org/abs/1911.04503} {arXiv:1911.04503
  [hep-th]} \BibitemShut {NoStop}%
\bibitem [{\citenamefont {Ruzziconi}\ and\ \citenamefont
  {Zwikel}(2026{\natexlab{a}})}]{Ruzziconi:2025fct}%
  \BibitemOpen
  \bibfield  {author} {\bibinfo {author} {\bibfnamefont {R.}~\bibnamefont
  {Ruzziconi}}\ and\ \bibinfo {author} {\bibfnamefont {C.}~\bibnamefont
  {Zwikel}},\ }\bibfield  {title} {\bibinfo {title} {{Celestial symmetries of
  black hole horizons}},\ }\href {https://doi.org/10.1103/gx7p-8k34} {\bibfield
   {journal} {\bibinfo  {journal} {Phys. Rev. D}\ }\textbf {\bibinfo {volume}
  {113}},\ \bibinfo {pages} {L041504} (\bibinfo {year} {2026}{\natexlab{a}})},\
  \Eprint {https://arxiv.org/abs/2504.08027} {arXiv:2504.08027 [hep-th]}
  \BibitemShut {NoStop}%
\bibitem [{\citenamefont {Agrawal}\ \emph {et~al.}(2026)\citenamefont
  {Agrawal}, \citenamefont {Charalambous},\ and\ \citenamefont
  {Donnay}}]{Agrawal:2025fsv}%
  \BibitemOpen
  \bibfield  {author} {\bibinfo {author} {\bibfnamefont {S.}~\bibnamefont
  {Agrawal}}, \bibinfo {author} {\bibfnamefont {P.}~\bibnamefont
  {Charalambous}},\ and\ \bibinfo {author} {\bibfnamefont {L.}~\bibnamefont
  {Donnay}},\ }\bibfield  {title} {\bibinfo {title} {{Null infinity as an
  inverted extremal horizon: Matching an infinite set of conserved quantities
  for gravitational perturbations}},\ }\href
  {https://doi.org/10.21468/SciPostPhys.20.2.054} {\bibfield  {journal}
  {\bibinfo  {journal} {SciPost Phys.}\ }\textbf {\bibinfo {volume} {20}},\
  \bibinfo {pages} {054} (\bibinfo {year} {2026})},\ \Eprint
  {https://arxiv.org/abs/2506.15526} {arXiv:2506.15526 [hep-th]} \BibitemShut
  {NoStop}%
\bibitem [{\citenamefont {Ruzziconi}\ and\ \citenamefont
  {Zwikel}(2026{\natexlab{b}})}]{Ruzziconi:2025fuy}%
  \BibitemOpen
  \bibfield  {author} {\bibinfo {author} {\bibfnamefont {R.}~\bibnamefont
  {Ruzziconi}}\ and\ \bibinfo {author} {\bibfnamefont {C.}~\bibnamefont
  {Zwikel}},\ }\bibfield  {title} {\bibinfo {title} {{Celestial
  $Lw_{1+\infty}$ symmetries and subleading phase space of null
  hypersurfaces}},\ }\href {https://doi.org/10.1103/hrbd-cmr7} {\bibfield
  {journal} {\bibinfo  {journal} {Phys. Rev. D}\ }\textbf {\bibinfo {volume}
  {113}},\ \bibinfo {pages} {044067} (\bibinfo {year} {2026}{\natexlab{b}})},\
  \Eprint {https://arxiv.org/abs/2511.07525} {arXiv:2511.07525 [hep-th]}
  \BibitemShut {NoStop}%
\bibitem [{\citenamefont {{Fletcher}}\ and\ \citenamefont
  {{Lun}}(1996)}]{Fletcher-Lun}%
  \BibitemOpen
  \bibfield  {author} {\bibinfo {author} {\bibfnamefont {S.~J.}\ \bibnamefont
  {{Fletcher}}}\ and\ \bibinfo {author} {\bibfnamefont {A.~W.~C.}\ \bibnamefont
  {{Lun}}},\ }\bibfield  {title} {\bibinfo {title} {{Bondi-Sachs metrics and
  exact solutions}},\ }in\ \href@noop {} {\emph {\bibinfo {booktitle} {{7th
  Marcel Grossman Meeting on General Relativity}}}},\ \bibinfo {editor} {edited
  by\ \bibinfo {editor} {\bibfnamefont {R.~T.}\ \bibnamefont {Jantzen}}\ and\
  \bibinfo {editor} {\bibfnamefont {G.~M.}\ \bibnamefont {Keiser}}}\ (\bibinfo
  {publisher} {World Scientific, Singapore},\ \bibinfo {year} {1996})\ p.\
  \bibinfo {pages} {296–298}\BibitemShut {NoStop}%
\bibitem [{\citenamefont {{Fletcher}}\ and\ \citenamefont
  {{Lun}}(2003)}]{Fletcher-Lun2003}%
  \BibitemOpen
  \bibfield  {author} {\bibinfo {author} {\bibfnamefont {S.~J.}\ \bibnamefont
  {{Fletcher}}}\ and\ \bibinfo {author} {\bibfnamefont {A.~W.~C.}\ \bibnamefont
  {{Lun}}},\ }\bibfield  {title} {\bibinfo {title} {{The Kerr spacetime in
  generalized Bondi Sachs coordinates}},\ }\href
  {https://doi.org/10.1088/0264-9381/20/19/302} {\bibfield  {journal} {\bibinfo
   {journal} {Classical and Quantum Gravity}\ }\textbf {\bibinfo {volume}
  {20}},\ \bibinfo {pages} {4153} (\bibinfo {year} {2003})}\BibitemShut
  {NoStop}%
\bibitem [{\citenamefont {Venter}\ and\ \citenamefont
  {Bishop}(2006)}]{Venter:2005cs}%
  \BibitemOpen
  \bibfield  {author} {\bibinfo {author} {\bibfnamefont {L.~R.}\ \bibnamefont
  {Venter}}\ and\ \bibinfo {author} {\bibfnamefont {N.~T.}\ \bibnamefont
  {Bishop}},\ }\bibfield  {title} {\bibinfo {title} {{Numerical validation of
  the Kerr metric in Bondi-Sachs form}},\ }\href
  {https://doi.org/10.1103/PhysRevD.73.084023} {\bibfield  {journal} {\bibinfo
  {journal} {Phys. Rev. D}\ }\textbf {\bibinfo {volume} {73}},\ \bibinfo
  {pages} {084023} (\bibinfo {year} {2006})},\ \Eprint
  {https://arxiv.org/abs/gr-qc/0506077} {arXiv:gr-qc/0506077} \BibitemShut
  {NoStop}%
\bibitem [{\citenamefont {Evans}(2010)}]{Evans}%
  \BibitemOpen
  \bibfield  {author} {\bibinfo {author} {\bibfnamefont {C.}~\bibnamefont
  {Evans}},\ }\bibfield  {title} {\bibinfo {title} {{Nonlinear first-order
  PDE}},\ }in\ \href@noop {} {\emph {\bibinfo {booktitle} {{Partial
  Differential Equations: Second Edition}}}}\ (\bibinfo  {publisher} {American
  Mathematical Society, Providence, Rhode Island},\ \bibinfo {year}
  {2010})\BibitemShut {NoStop}%
\bibitem [{\citenamefont {Levandosky}(2002)}]{Levandosky}%
  \BibitemOpen
  \bibfield  {author} {\bibinfo {author} {\bibfnamefont {J.}~\bibnamefont
  {Levandosky}},\ }\bibfield  {title} {\bibinfo {title} {First-order equations:
  Method of characteristics}\ }(\bibinfo {year} {2002})\ \bibinfo {note}
  {\url{https://web.stanford.edu/class/math220a/handouts/firstorder.pdf}}\BibitemShut
  {NoStop}%
\bibitem [{\citenamefont {He}\ and\ \citenamefont {Mitra}(2019)}]{He:2019jjk}%
  \BibitemOpen
  \bibfield  {author} {\bibinfo {author} {\bibfnamefont {T.}~\bibnamefont
  {He}}\ and\ \bibinfo {author} {\bibfnamefont {P.}~\bibnamefont {Mitra}},\
  }\bibfield  {title} {\bibinfo {title} {{Asymptotic symmetries and
  Weinberg{\textquoteright}s soft photon theorem in Mink$_{d+2}$}},\ }\href
  {https://doi.org/10.1007/JHEP10(2019)213} {\bibfield  {journal} {\bibinfo
  {journal} {JHEP}\ }\textbf {\bibinfo {volume} {10}},\ \bibinfo {pages}
  {213} (2019)},\ \Eprint {https://arxiv.org/abs/1903.02608} {arXiv:1903.02608
  [hep-th]} \BibitemShut {NoStop}%
\bibitem [{compere()}]{compere}%
  \BibitemOpen
  \href@noop {} {}\bibinfo {note} {In \cite{Compere:2016jwb}, a radial shift $r=\bar r - \p_\zeta \p_{\bar \zeta} f_0$ was introduced, which corresponds to the integration constant in $\frac{\td r}{\td s}=1$. However, this shift does not affect the definition of the supertranslation.}\BibitemShut {Stop}%
\bibitem [{\citenamefont {Frauendiener}(1992)}]{Frauendiener}%
  \BibitemOpen
  \bibfield  {author} {\bibinfo {author} {\bibfnamefont {J.}~\bibnamefont
  {Frauendiener}},\ }\bibfield  {title} {\bibinfo {title} {{Note on the memory
  effect}},\ }\href {https://doi.org/10.1088/0264-9381/9/6/018} {\bibfield
  {journal} {\bibinfo  {journal} {Class. Quant. Grav.}\ }\textbf {\bibinfo
  {volume} {9}},\ \bibinfo {pages} {1639} (\bibinfo {year} {1992})}\BibitemShut
  {NoStop}%
\bibitem [{\citenamefont {Steinbauer}(1998)}]{Steinbauer:1997dw}%
  \BibitemOpen
  \bibfield  {author} {\bibinfo {author} {\bibfnamefont {R.}~\bibnamefont
  {Steinbauer}},\ }\bibfield  {title} {\bibinfo {title} {{Geodesics and
  geodesic deviation for impulsive gravitational waves}},\ }\href
  {https://doi.org/10.1063/1.532283} {\bibfield  {journal} {\bibinfo  {journal}
  {J. Math. Phys.}\ }\textbf {\bibinfo {volume} {39}},\ \bibinfo {pages} {2201}
  (\bibinfo {year} {1998})},\ \Eprint {https://arxiv.org/abs/gr-qc/9710119}
  {arXiv:gr-qc/9710119} \BibitemShut {NoStop}%
\bibitem [{\citenamefont {O'Loughlin}\ and\ \citenamefont
  {Demirchian}(2019)}]{OLoughlin:2018ebk}%
  \BibitemOpen
  \bibfield  {author} {\bibinfo {author} {\bibfnamefont {M.}~\bibnamefont
  {O'Loughlin}}\ and\ \bibinfo {author} {\bibfnamefont {H.}~\bibnamefont
  {Demirchian}},\ }\bibfield  {title} {\bibinfo {title} {{Geodesic congruences,
  impulsive gravitational waves and gravitational memory}},\ }\href
  {https://doi.org/10.1103/PhysRevD.99.024031} {\bibfield  {journal} {\bibinfo
  {journal} {Phys. Rev. D}\ }\textbf {\bibinfo {volume} {99}},\ \bibinfo
  {pages} {024031} (\bibinfo {year} {2019})},\ \Eprint
  {https://arxiv.org/abs/1808.04886} {arXiv:1808.04886 [hep-th]} \BibitemShut
  {NoStop}%
\bibitem [{\citenamefont {Steinbauer}(2019)}]{Steinbauer:2018iis}%
  \BibitemOpen
  \bibfield  {author} {\bibinfo {author} {\bibfnamefont {R.}~\bibnamefont
  {Steinbauer}},\ }\bibfield  {title} {\bibinfo {title} {{The memory effect in
  impulsive plane waves: comments, corrections, clarifications}},\ }\href
  {https://doi.org/10.1088/1361-6382/ab127d} {\bibfield  {journal} {\bibinfo
  {journal} {Class. Quant. Grav.}\ }\textbf {\bibinfo {volume} {36}},\ \bibinfo
  {pages} {098001} (\bibinfo {year} {2019})},\ \Eprint
  {https://arxiv.org/abs/1811.10940} {arXiv:1811.10940 [gr-qc]} \BibitemShut
  {NoStop}%
\bibitem [{\citenamefont {Bhattacharjee}\ \emph {et~al.}(2019)\citenamefont
  {Bhattacharjee}, \citenamefont {Kumar},\ and\ \citenamefont
  {Bhattacharyya}}]{Bhattacharjee:2019jaf}%
  \BibitemOpen
  \bibfield  {author} {\bibinfo {author} {\bibfnamefont {S.}~\bibnamefont
  {Bhattacharjee}}, \bibinfo {author} {\bibfnamefont {S.}~\bibnamefont
  {Kumar}},\ and\ \bibinfo {author} {\bibfnamefont {A.}~\bibnamefont
  {Bhattacharyya}},\ }\bibfield  {title} {\bibinfo {title} {{Memory Effect and
  BMS-like Symmetries for Impulsive Gravitational Waves}},\ }\href
  {https://doi.org/10.1103/PhysRevD.100.084010} {\bibfield  {journal} {\bibinfo
   {journal} {Phys. Rev. D}\ }\textbf {\bibinfo {volume} {100}},\ \bibinfo
  {pages} {084010} (\bibinfo {year} {2019})},\ \Eprint
  {https://arxiv.org/abs/1905.12905} {arXiv:1905.12905 [hep-th]} \BibitemShut
  {NoStop}%
\bibitem [{\citenamefont {Blau}\ and\ \citenamefont
  {O'Loughlin}(2016{\natexlab{a}})}]{Blau:2015nee}%
  \BibitemOpen
  \bibfield  {author} {\bibinfo {author} {\bibfnamefont {M.}~\bibnamefont
  {Blau}}\ and\ \bibinfo {author} {\bibfnamefont {M.}~\bibnamefont
  {O'Loughlin}},\ }\bibfield  {title} {\bibinfo {title} {{Horizon Shells and
  BMS-like Soldering Transformations}},\ }\href
  {https://doi.org/10.1007/JHEP03(2016)029} {\bibfield  {journal} {\bibinfo
  {journal} {JHEP}\ }\textbf {\bibinfo {volume} {03}},\ \bibinfo {pages}
  {029} (2016)},\ \Eprint {https://arxiv.org/abs/1512.02858} {arXiv:1512.02858
  [hep-th]} \BibitemShut {NoStop}%
\bibitem [{\citenamefont {Blau}\ and\ \citenamefont
  {O'Loughlin}(2016{\natexlab{b}})}]{Blau:2016juv}%
  \BibitemOpen
  \bibfield  {author} {\bibinfo {author} {\bibfnamefont {M.}~\bibnamefont
  {Blau}}\ and\ \bibinfo {author} {\bibfnamefont {M.}~\bibnamefont
  {O'Loughlin}},\ }\bibfield  {title} {\bibinfo {title} {{Horizon Shells:
  Classical Structure at the Horizon of a Black Hole}},\ }\href
  {https://doi.org/10.1142/S0218271816440107} {\bibfield  {journal} {\bibinfo
  {journal} {Int. J. Mod. Phys. D}\ }\textbf {\bibinfo {volume} {25}},\
  \bibinfo {pages} {1644010} (\bibinfo {year} {2016}{\natexlab{b}})},\ \Eprint
  {https://arxiv.org/abs/1604.01181} {arXiv:1604.01181 [hep-th]} \BibitemShut
  {NoStop}%
\bibitem [{\citenamefont {Bhattacharjee}\ and\ \citenamefont
  {Bhattacharyya}(2018)}]{Bhattacharjee:2017gkh}%
  \BibitemOpen
  \bibfield  {author} {\bibinfo {author} {\bibfnamefont {S.}~\bibnamefont
  {Bhattacharjee}}\ and\ \bibinfo {author} {\bibfnamefont {A.}~\bibnamefont
  {Bhattacharyya}},\ }\bibfield  {title} {\bibinfo {title} {{Soldering freedom
  and Bondi-Metzner-Sachs-like transformations}},\ }\href
  {https://doi.org/10.1103/PhysRevD.98.104009} {\bibfield  {journal} {\bibinfo
  {journal} {Phys. Rev. D}\ }\textbf {\bibinfo {volume} {98}},\ \bibinfo
  {pages} {104009} (\bibinfo {year} {2018})},\ \Eprint
  {https://arxiv.org/abs/1707.01112} {arXiv:1707.01112 [hep-th]} \BibitemShut
  {NoStop}%
\bibitem [{\citenamefont {{Labeyrie}}(1993)}]{Labeyrie}%
  \BibitemOpen
  \bibfield  {author} {\bibinfo {author} {\bibfnamefont {A.}~\bibnamefont
  {{Labeyrie}}},\ }\bibfield  {title} {\bibinfo {title} {{Lensing effects of
  gravitational radiation near celestial sources}},\ }\href@noop {} {\bibfield
   {journal} {\bibinfo  {journal} {Astron. Astrophys.}\ }\textbf
  {\bibinfo {volume} {268}},\ \bibinfo {pages} {823} (\bibinfo {year}
  {1993})}\BibitemShut {NoStop}%
\bibitem [{\citenamefont {Durrer}(1994)}]{Durrer:1994uu}%
  \BibitemOpen
  \bibfield  {author} {\bibinfo {author} {\bibfnamefont {R.}~\bibnamefont
  {Durrer}},\ }\bibfield  {title} {\bibinfo {title} {{Light deflection in
  perturbed Friedmann universes}},\ }\href
  {https://doi.org/10.1103/PhysRevLett.72.3301} {\bibfield  {journal} {\bibinfo
   {journal} {Phys. Rev. Lett.}\ }\textbf {\bibinfo {volume} {72}},\ \bibinfo
  {pages} {3301} (\bibinfo {year} {1994})},\ \Eprint
  {https://arxiv.org/abs/astro-ph/9401033} {arXiv:astro-ph/9401033}
  \BibitemShut {NoStop}%
\bibitem [{\citenamefont {Kaiser}\ and\ \citenamefont
  {Jaffe}(1997)}]{Kaiser:1996wk}%
  \BibitemOpen
  \bibfield  {author} {\bibinfo {author} {\bibfnamefont {N.}~\bibnamefont
  {Kaiser}}\ and\ \bibinfo {author} {\bibfnamefont {A.~H.}\ \bibnamefont
  {Jaffe}},\ }\bibfield  {title} {\bibinfo {title} {{Bending of light by
  gravity waves}},\ }\href {https://doi.org/10.1086/304357} {\bibfield
  {journal} {\bibinfo  {journal} {Astrophys. J.}\ }\textbf {\bibinfo {volume}
  {484}},\ \bibinfo {pages} {545} (\bibinfo {year} {1997})},\ \Eprint
  {https://arxiv.org/abs/astro-ph/9609043} {arXiv:astro-ph/9609043}
  \BibitemShut {NoStop}%
\bibitem [{\citenamefont {Damour}\ and\ \citenamefont
  {Esposito-Farese}(1998)}]{Damour:1998jm}%
  \BibitemOpen
  \bibfield  {author} {\bibinfo {author} {\bibfnamefont {T.}~\bibnamefont
  {Damour}}\ and\ \bibinfo {author} {\bibfnamefont {G.}~\bibnamefont
  {Esposito-Farese}},\ }\bibfield  {title} {\bibinfo {title} {{Light deflection
  by gravitational waves from localized sources}},\ }\href
  {https://doi.org/10.1103/PhysRevD.58.044003} {\bibfield  {journal} {\bibinfo
  {journal} {Phys. Rev. D}\ }\textbf {\bibinfo {volume} {58}},\ \bibinfo
  {pages} {044003} (\bibinfo {year} {1998})},\ \Eprint
  {https://arxiv.org/abs/gr-qc/9802019} {arXiv:gr-qc/9802019} \BibitemShut
  {NoStop}%
\bibitem [{\citenamefont {Zhong}\ \emph {et~al.}(2025)\citenamefont {Zhong},
  \citenamefont {Cardoso},\ and\ \citenamefont {Chen}}]{Zhong:2024ysg}%
  \BibitemOpen
  \bibfield  {author} {\bibinfo {author} {\bibfnamefont {Z.}~\bibnamefont
  {Zhong}}, \bibinfo {author} {\bibfnamefont {V.}~\bibnamefont {Cardoso}},\
  and\ \bibinfo {author} {\bibfnamefont {Y.}~\bibnamefont {Chen}},\ }\bibfield
  {title} {\bibinfo {title} {{Dynamical Lensing Tomography of Black Hole
  Ringdowns}},\ }\href {https://doi.org/10.1103/PhysRevLett.134.211402}
  {\bibfield  {journal} {\bibinfo  {journal} {Phys. Rev. Lett.}\ }\textbf
  {\bibinfo {volume} {134}},\ \bibinfo {pages} {211402} (\bibinfo {year}
  {2025})},\ \Eprint {https://arxiv.org/abs/2408.10303} {arXiv:2408.10303
  [gr-qc]} \BibitemShut {NoStop}%
\bibitem [{displacement()}]{displacement}%
  \BibitemOpen
  \href@noop {} {}\bibinfo {note} {\blue{The well-known displacement memory effect corresponds to geodesic deviation induced by gravitational waves. In the gravitational lensing picture, the full trajectories of light rays encode the physical information about the gravitational memory effect.} }\BibitemShut {Stop}%
\bibitem [{\citenamefont {mao}()}]{mao}%
  \BibitemOpen
  \bibfield  {author} {\bibinfo {author} {\bibfnamefont {P.}\ \bibnamefont
  {Mao}},\ }\bibfield  {title} {\bibinfo {title} {{Supertranslation and null hypersurface quantization}},\ } {work in progress (2026)}\BibitemShut {NoStop}%
\bibitem [{\citenamefont {Bengtsson}\ \emph {et~al.}(1983)\citenamefont
  {Bengtsson}, \citenamefont {Bengtsson},\ and\ \citenamefont
  {Brink}}]{Bengtsson:1983pd}%
  \BibitemOpen
  \bibfield  {author} {\bibinfo {author} {\bibfnamefont {A.~K.~H.}\
  \bibnamefont {Bengtsson}}, \bibinfo {author} {\bibfnamefont {I.}~\bibnamefont
  {Bengtsson}},\ and\ \bibinfo {author} {\bibfnamefont {L.}~\bibnamefont
  {Brink}},\ }\bibfield  {title} {\bibinfo {title} {{Cubic Interaction Terms
  for Arbitrary Spin}},\ }\href {https://doi.org/10.1016/0550-3213(83)90140-2}
  {\bibfield  {journal} {\bibinfo  {journal} {Nucl. Phys. B}\ }\textbf
  {\bibinfo {volume} {227}},\ \bibinfo {pages} {31} (\bibinfo {year}
  {1983})}\BibitemShut {NoStop}%
\bibitem [{\citenamefont {Ananth}\ \emph {et~al.}(2006)\citenamefont {Ananth},
  \citenamefont {Brink}, \citenamefont {Heise},\ and\ \citenamefont
  {Svendsen}}]{Ananth:2006fh}%
  \BibitemOpen
  \bibfield  {author} {\bibinfo {author} {\bibfnamefont {S.}~\bibnamefont
  {Ananth}}, \bibinfo {author} {\bibfnamefont {L.}~\bibnamefont {Brink}},
  \bibinfo {author} {\bibfnamefont {R.}~\bibnamefont {Heise}},\ and\ \bibinfo
  {author} {\bibfnamefont {H.~G.}\ \bibnamefont {Svendsen}},\ }\bibfield
  {title} {\bibinfo {title} {{The N=8 Supergravity Hamiltonian as a Quadratic
  Form}},\ }\href {https://doi.org/10.1016/j.nuclphysb.2006.07.014} {\bibfield
  {journal} {\bibinfo  {journal} {Nucl. Phys. B}\ }\textbf {\bibinfo {volume}
  {753}},\ \bibinfo {pages} {195} (\bibinfo {year} {2006})},\ \Eprint
  {https://arxiv.org/abs/hep-th/0607019} {arXiv:hep-th/0607019} \BibitemShut
  {NoStop}%
\bibitem [{\citenamefont {Brown}\ and\ \citenamefont
  {York}(1993)}]{Brown:1992br}%
  \BibitemOpen
  \bibfield  {author} {\bibinfo {author} {\bibfnamefont {J.~D.}\ \bibnamefont
  {Brown}}\ and\ \bibinfo {author} {\bibfnamefont {J.~W.}\ \bibnamefont {York},
  \bibfnamefont {Jr.}},\ }\bibfield  {title} {\bibinfo {title} {{Quasilocal
  energy and conserved charges derived from the gravitational action}},\ }\href
  {https://doi.org/10.1103/PhysRevD.47.1407} {\bibfield  {journal} {\bibinfo
  {journal} {Phys. Rev. D}\ }\textbf {\bibinfo {volume} {47}},\ \bibinfo
  {pages} {1407} (\bibinfo {year} {1993})},\ \Eprint
  {https://arxiv.org/abs/gr-qc/9209012} {arXiv:gr-qc/9209012} \BibitemShut
  {NoStop}%
\bibitem [{\citenamefont {Parattu}\ \emph {et~al.}(2016)\citenamefont
  {Parattu}, \citenamefont {Chakraborty},\ and\ \citenamefont
  {Padmanabhan}}]{Parattu:2016trq}%
  \BibitemOpen
  \bibfield  {author} {\bibinfo {author} {\bibfnamefont {K.}~\bibnamefont
  {Parattu}}, \bibinfo {author} {\bibfnamefont {S.}~\bibnamefont
  {Chakraborty}},\ and\ \bibinfo {author} {\bibfnamefont {T.}~\bibnamefont
  {Padmanabhan}},\ }\bibfield  {title} {\bibinfo {title} {{Variational
  Principle for Gravity with Null and Non-null boundaries: A Unified Boundary
  Counter-term}},\ }\href {https://doi.org/10.1140/epjc/s10052-016-3979-y}
  {\bibfield  {journal} {\bibinfo  {journal} {Eur. Phys. J. C}\ }\textbf
  {\bibinfo {volume} {76}},\ \bibinfo {pages} {129} (\bibinfo {year} {2016})},\
  \Eprint {https://arxiv.org/abs/1602.07546} {arXiv:1602.07546 [gr-qc]}
  \BibitemShut {NoStop}%
\bibitem [{\citenamefont {Donnelly}\ and\ \citenamefont
  {Giddings}(2016)}]{Donnelly:2016rvo}%
  \BibitemOpen
  \bibfield  {author} {\bibinfo {author} {\bibfnamefont {W.}~\bibnamefont
  {Donnelly}}\ and\ \bibinfo {author} {\bibfnamefont {S.~B.}\ \bibnamefont
  {Giddings}},\ }\bibfield  {title} {\bibinfo {title} {{Observables,
  gravitational dressing, and obstructions to locality and subsystems}},\
  }\href {https://doi.org/10.1103/PhysRevD.94.104038} {\bibfield  {journal}
  {\bibinfo  {journal} {Phys. Rev. D}\ }\textbf {\bibinfo {volume} {94}},\
  \bibinfo {pages} {104038} (\bibinfo {year} {2016})},\ \Eprint
  {https://arxiv.org/abs/1607.01025} {arXiv:1607.01025 [hep-th]} \BibitemShut
  {NoStop}%
\bibitem [{\citenamefont {Smirnov}\ and\ \citenamefont
  {Zamolodchikov}(2017)}]{Smirnov:2016lqw}%
  \BibitemOpen
  \bibfield  {author} {\bibinfo {author} {\bibfnamefont {F.~A.}\ \bibnamefont
  {Smirnov}}\ and\ \bibinfo {author} {\bibfnamefont {A.~B.}\ \bibnamefont
  {Zamolodchikov}},\ }\bibfield  {title} {\bibinfo {title} {{On space of
  integrable quantum field theories}},\ }\href
  {https://doi.org/10.1016/j.nuclphysb.2016.12.014} {\bibfield  {journal}
  {\bibinfo  {journal} {Nucl. Phys. B}\ }\textbf {\bibinfo {volume} {915}},\
  \bibinfo {pages} {363} (\bibinfo {year} {2017})},\ \Eprint
  {https://arxiv.org/abs/1608.05499} {arXiv:1608.05499 [hep-th]} \BibitemShut
  {NoStop}%
\bibitem [{\citenamefont {Cavagli{\`a}}\ \emph {et~al.}(2016)\citenamefont
  {Cavagli{\`a}}, \citenamefont {Negro}, \citenamefont {Sz{\'e}cs{\'e}nyi},\
  and\ \citenamefont {Tateo}}]{Cavaglia:2016oda}%
  \BibitemOpen
  \bibfield  {author} {\bibinfo {author} {\bibfnamefont {A.}~\bibnamefont
  {Cavagli{\`a}}}, \bibinfo {author} {\bibfnamefont {S.}~\bibnamefont {Negro}},
  \bibinfo {author} {\bibfnamefont {I.~M.}\ \bibnamefont {Sz{\'e}cs{\'e}nyi}},\
  and\ \bibinfo {author} {\bibfnamefont {R.}~\bibnamefont {Tateo}},\ }\bibfield
   {title} {\bibinfo {title} {{$T \bar{T}$-deformed 2D Quantum Field
  Theories}},\ }\href {https://doi.org/10.1007/JHEP10(2016)112} {\bibfield
  {journal} {\bibinfo  {journal} {JHEP}\ }\textbf {\bibinfo {volume} {10}},\
  \bibinfo {pages} {112} (2016)},\ \Eprint {https://arxiv.org/abs/1608.05534}
  {arXiv:1608.05534 [hep-th]} \BibitemShut {NoStop}%
\bibitem [{\citenamefont {Jiang}(2021)}]{Jiang:2019epa}%
  \BibitemOpen
  \bibfield  {author} {\bibinfo {author} {\bibfnamefont {Y.}~\bibnamefont
  {Jiang}},\ }\bibfield  {title} {\bibinfo {title} {{A pedagogical review on
  solvable irrelevant deformations of 2D quantum field theory}},\ }\href
  {https://doi.org/10.1088/1572-9494/abe4c9} {\bibfield  {journal} {\bibinfo
  {journal} {Commun. Theor. Phys.}\ }\textbf {\bibinfo {volume} {73}},\
  \bibinfo {pages} {057201} (\bibinfo {year} {2021})},\ \Eprint
  {https://arxiv.org/abs/1904.13376} {arXiv:1904.13376 [hep-th]} \BibitemShut
  {NoStop}%
\bibitem [{\citenamefont {He}\ \emph {et~al.}(2025)\citenamefont {He},
  \citenamefont {Li}, \citenamefont {Ouyang},\ and\ \citenamefont
  {Sun}}]{He:2025ppz}%
  \BibitemOpen
  \bibfield  {author} {\bibinfo {author} {\bibfnamefont {S.}~\bibnamefont
  {He}}, \bibinfo {author} {\bibfnamefont {Y.}~\bibnamefont {Li}}, \bibinfo
  {author} {\bibfnamefont {H.}~\bibnamefont {Ouyang}},\ and\ \bibinfo {author}
  {\bibfnamefont {Y.}~\bibnamefont {Sun}},\ }\bibfield  {title} {\bibinfo
  {title} {{$T\overline{T}$ deformation: Introduction and some recent
  advances}},\ }\href {https://doi.org/10.1007/s11433-025-2708-2} {\bibfield
  {journal} {\bibinfo  {journal} {Sci. China Phys. Mech. Astron.}\ }\textbf
  {\bibinfo {volume} {68}},\ \bibinfo {pages} {101001} (\bibinfo {year}
  {2025})},\ \Eprint {https://arxiv.org/abs/2503.09997} {arXiv:2503.09997
  [hep-th]} \BibitemShut {NoStop}%
\bibitem [{\citenamefont {McGough}\ \emph {et~al.}(2018)\citenamefont
  {McGough}, \citenamefont {Mezei},\ and\ \citenamefont
  {Verlinde}}]{McGough:2016lol}%
  \BibitemOpen
  \bibfield  {author} {\bibinfo {author} {\bibfnamefont {L.}~\bibnamefont
  {McGough}}, \bibinfo {author} {\bibfnamefont {M.}~\bibnamefont {Mezei}},\
  and\ \bibinfo {author} {\bibfnamefont {H.}~\bibnamefont {Verlinde}},\
  }\bibfield  {title} {\bibinfo {title} {{Moving the CFT into the bulk with $
  T\overline{T} $}},\ }\href {https://doi.org/10.1007/JHEP04(2018)010}
  {\bibfield  {journal} {\bibinfo  {journal} {JHEP}\ }\textbf {\bibinfo
  {volume} {04}},\ \bibinfo {pages} {010} (2018)},\ \Eprint
  {https://arxiv.org/abs/1611.03470} {arXiv:1611.03470 [hep-th]} \BibitemShut
  {NoStop}%
\bibitem [{\citenamefont {Kraus}\ \emph {et~al.}(2018)\citenamefont {Kraus},
  \citenamefont {Liu},\ and\ \citenamefont {Marolf}}]{Kraus:2018xrn}%
  \BibitemOpen
  \bibfield  {author} {\bibinfo {author} {\bibfnamefont {P.}~\bibnamefont
  {Kraus}}, \bibinfo {author} {\bibfnamefont {J.}~\bibnamefont {Liu}},\ and\
  \bibinfo {author} {\bibfnamefont {D.}~\bibnamefont {Marolf}},\ }\bibfield
  {title} {\bibinfo {title} {{Cutoff AdS$_{3}$ versus the $ T\overline{T} $
  deformation}},\ }\href {https://doi.org/10.1007/JHEP07(2018)027} {\bibfield
  {journal} {\bibinfo  {journal} {JHEP}\ }\textbf {\bibinfo {volume} {07}},\
  \bibinfo {pages} {027} (2018)},\ \Eprint {https://arxiv.org/abs/1801.02714}
  {arXiv:1801.02714 [hep-th]} \BibitemShut {NoStop}%
\bibitem [{\citenamefont {Hartman}\ \emph {et~al.}(2019)\citenamefont
  {Hartman}, \citenamefont {Kruthoff}, \citenamefont {Shaghoulian},\ and\
  \citenamefont {Tajdini}}]{Hartman:2018tkw}%
  \BibitemOpen
  \bibfield  {author} {\bibinfo {author} {\bibfnamefont {T.}~\bibnamefont
  {Hartman}}, \bibinfo {author} {\bibfnamefont {J.}~\bibnamefont {Kruthoff}},
  \bibinfo {author} {\bibfnamefont {E.}~\bibnamefont {Shaghoulian}},\ and\
  \bibinfo {author} {\bibfnamefont {A.}~\bibnamefont {Tajdini}},\ }\bibfield
  {title} {\bibinfo {title} {{Holography at finite cutoff with a $T^2$
  deformation}},\ }\href {https://doi.org/10.1007/JHEP03(2019)004} {\bibfield
  {journal} {\bibinfo  {journal} {JHEP}\ }\textbf {\bibinfo {volume} {03}},\
  \bibinfo {pages} {004} (2019)},\ \Eprint {https://arxiv.org/abs/1807.11401}
  {arXiv:1807.11401 [hep-th]} \BibitemShut {NoStop}%
\bibitem [{\citenamefont {Guica}\ and\ \citenamefont
  {Monten}(2021)}]{Guica:2019nzm}%
  \BibitemOpen
  \bibfield  {author} {\bibinfo {author} {\bibfnamefont {M.}~\bibnamefont
  {Guica}}\ and\ \bibinfo {author} {\bibfnamefont {R.}~\bibnamefont {Monten}},\
  }\bibfield  {title} {\bibinfo {title} {{$T\bar T$ and the mirage of a bulk
  cutoff}},\ }\href {https://doi.org/10.21468/SciPostPhys.10.2.024} {\bibfield
  {journal} {\bibinfo  {journal} {SciPost Phys.}\ }\textbf {\bibinfo {volume}
  {10}},\ \bibinfo {pages} {024} (\bibinfo {year} {2021})},\ \Eprint
  {https://arxiv.org/abs/1906.11251} {arXiv:1906.11251 [hep-th]} \BibitemShut
  {NoStop}%
\bibitem [{\citenamefont {Jafari}\ \emph {et~al.}(2020)\citenamefont {Jafari},
  \citenamefont {Naseh},\ and\ \citenamefont {Zolfi}}]{Jafari:2019qns}%
  \BibitemOpen
  \bibfield  {author} {\bibinfo {author} {\bibfnamefont {G.}~\bibnamefont
  {Jafari}}, \bibinfo {author} {\bibfnamefont {A.}~\bibnamefont {Naseh}},\ and\
  \bibinfo {author} {\bibfnamefont {H.}~\bibnamefont {Zolfi}},\ }\bibfield
  {title} {\bibinfo {title} {{Path Integral Optimization for $T\bar{T}$
  Deformation}},\ }\href {https://doi.org/10.1103/PhysRevD.101.026007}
  {\bibfield  {journal} {\bibinfo  {journal} {Phys. Rev. D}\ }\textbf {\bibinfo
  {volume} {101}},\ \bibinfo {pages} {026007} (\bibinfo {year} {2020})},\
  \Eprint {https://arxiv.org/abs/1909.02357} {arXiv:1909.02357 [hep-th]}
  \BibitemShut {NoStop}%
\bibitem [{\citenamefont {He}\ \emph {et~al.}(2023)\citenamefont {He},
  \citenamefont {Li}, \citenamefont {Li},\ and\ \citenamefont
  {Zhang}}]{He:2023hoj}%
  \BibitemOpen
  \bibfield  {author} {\bibinfo {author} {\bibfnamefont {S.}~\bibnamefont
  {He}}, \bibinfo {author} {\bibfnamefont {Y.}~\bibnamefont {Li}}, \bibinfo
  {author} {\bibfnamefont {Y.-Z.}\ \bibnamefont {Li}},\ and\ \bibinfo {author}
  {\bibfnamefont {Y.}~\bibnamefont {Zhang}},\ }\bibfield  {title} {\bibinfo
  {title} {{Holographic torus correlators of stress tensor in
  AdS$_{3}$/CFT$_{2}$}},\ }\href {https://doi.org/10.1007/JHEP06(2023)116}
  {\bibfield  {journal} {\bibinfo  {journal} {JHEP}\ }\textbf {\bibinfo
  {volume} {06}},\ \bibinfo {pages} {116} (2023)},\ \Eprint
  {https://arxiv.org/abs/2303.13280} {arXiv:2303.13280 [hep-th]} \BibitemShut
  {NoStop}%
\bibitem [{\citenamefont {He}\ \emph {et~al.}(2024)\citenamefont {He},
  \citenamefont {Li},\ and\ \citenamefont {Zhang}}]{He:2023knl}%
  \BibitemOpen
  \bibfield  {author} {\bibinfo {author} {\bibfnamefont {S.}~\bibnamefont
  {He}}, \bibinfo {author} {\bibfnamefont {Y.-Z.}\ \bibnamefont {Li}},\ and\
  \bibinfo {author} {\bibfnamefont {Y.}~\bibnamefont {Zhang}},\ }\bibfield
  {title} {\bibinfo {title} {{Holographic torus correlators in AdS$_{3}$
  gravity coupled to scalar field}},\ }\href
  {https://doi.org/10.1007/JHEP05(2024)254} {\bibfield  {journal} {\bibinfo
  {journal} {JHEP}\ }\textbf {\bibinfo {volume} {05}},\ \bibinfo {pages}
  {254} (2024)},\ \Eprint {https://arxiv.org/abs/2311.09636} {arXiv:2311.09636
  [hep-th]} \BibitemShut {NoStop}%
\bibitem [{moving()}]{moving}%
  \BibitemOpen
  \href@noop {} {}\bibinfo {note} {Moving boundaries from the viewpoint of covariant phase space formalism can be found, e.g., in \cite{Parvizi:2025shq,Parvizi:2025wsg,Sheikh-Jabbari:2025kjd,Adami:2025pqr}.}\BibitemShut {Stop}%
\bibitem [{\citenamefont {Parvizi}\ \emph
  {et~al.}(2025{\natexlab{a}})\citenamefont {Parvizi}, \citenamefont
  {Sheikh-Jabbari},\ and\ \citenamefont {Taghiloo}}]{Parvizi:2025shq}%
  \BibitemOpen
  \bibfield  {author} {\bibinfo {author} {\bibfnamefont {A.}~\bibnamefont
  {Parvizi}}, \bibinfo {author} {\bibfnamefont {M.~M.}\ \bibnamefont
  {Sheikh-Jabbari}},\ and\ \bibinfo {author} {\bibfnamefont {V.}~\bibnamefont
  {Taghiloo}},\ }\bibfield  {title} {\bibinfo {title} {{Freelance holography,
  part I: Setting boundary conditions free in gauge/gravity correspondence}},\
  }\href {https://doi.org/10.21468/SciPostPhys.19.2.043} {\bibfield  {journal}
  {\bibinfo  {journal} {SciPost Phys.}\ }\textbf {\bibinfo {volume} {19}},\
  \bibinfo {pages} {043} (\bibinfo {year} {2025}{\natexlab{a}})},\ \Eprint
  {https://arxiv.org/abs/2503.09371} {arXiv:2503.09371 [hep-th]} \BibitemShut
  {NoStop}%
\bibitem [{\citenamefont {Parvizi}\ \emph
  {et~al.}(2025{\natexlab{b}})\citenamefont {Parvizi}, \citenamefont
  {Sheikh-Jabbari},\ and\ \citenamefont {Taghiloo}}]{Parvizi:2025wsg}%
  \BibitemOpen
  \bibfield  {author} {\bibinfo {author} {\bibfnamefont {A.}~\bibnamefont
  {Parvizi}}, \bibinfo {author} {\bibfnamefont {M.~M.}\ \bibnamefont
  {Sheikh-Jabbari}},\ and\ \bibinfo {author} {\bibfnamefont {V.}~\bibnamefont
  {Taghiloo}},\ }\bibfield  {title} {\bibinfo {title} {{Freelance Holography,
  Part II: Moving Boundary in Gauge/Gravity Correspondence}},\ }\href
  {https://doi.org/10.21468/SciPostPhysCore.8.4.075} {\bibfield  {journal}
  {\bibinfo  {journal} {SciPost Phys. Core}\ }\textbf {\bibinfo {volume} {8}},\
  \bibinfo {pages} {075} (\bibinfo {year} {2025}{\natexlab{b}})},\ \Eprint
  {https://arxiv.org/abs/2503.09372} {arXiv:2503.09372 [hep-th]} \BibitemShut
  {NoStop}%
\bibitem [{\citenamefont {Sheikh-Jabbari}\ and\ \citenamefont
  {Taghiloo}(2025)}]{Sheikh-Jabbari:2025kjd}%
  \BibitemOpen
  \bibfield  {author} {\bibinfo {author} {\bibfnamefont {M.~M.}\ \bibnamefont
  {Sheikh-Jabbari}}\ and\ \bibinfo {author} {\bibfnamefont {V.}~\bibnamefont
  {Taghiloo}},\ }\bibfield  {title} {\bibinfo {title} {{AdS$_3$ Freelance
  Holography, A Detailed Analysis}},\ }\href@noop {} {\  (\bibinfo {year}
  {2025})},\ \Eprint {https://arxiv.org/abs/2510.10692} {arXiv:2510.10692
  [hep-th]} \BibitemShut {NoStop}%
\bibitem [{\citenamefont {Adami}\ \emph {et~al.}(2025)\citenamefont {Adami},
  \citenamefont {Sheikh-Jabbari},\ and\ \citenamefont
  {Taghiloo}}]{Adami:2025pqr}%
  \BibitemOpen
  \bibfield  {author} {\bibinfo {author} {\bibfnamefont {H.}~\bibnamefont
  {Adami}}, \bibinfo {author} {\bibfnamefont {M.~M.}\ \bibnamefont
  {Sheikh-Jabbari}},\ and\ \bibinfo {author} {\bibfnamefont {V.}~\bibnamefont
  {Taghiloo}},\ }\bibfield  {title} {\bibinfo {title} {{Gravity Is Induced By
  Renormalization Group Flow}},\ }\href@noop {} {\  (\bibinfo {year} {2025})},\
  \Eprint {https://arxiv.org/abs/2508.09633} {arXiv:2508.09633 [hep-th]}
  \BibitemShut {NoStop}%
\bibitem [{\citenamefont {Hou}(2023)}]{Hou:2022csf}%
  \BibitemOpen
  \bibfield  {author} {\bibinfo {author} {\bibfnamefont {J.}~\bibnamefont
  {Hou}},\ }\bibfield  {title} {\bibinfo {title} {{$ T\overline{T} $ flow as
  characteristic flows}},\ }\href {https://doi.org/10.1007/JHEP03(2023)243}
  {\bibfield  {journal} {\bibinfo  {journal} {JHEP}\ }\textbf {\bibinfo
  {volume} {03}},\ \bibinfo {pages} {243} (2023)},\ \Eprint
  {https://arxiv.org/abs/2208.05391} {arXiv:2208.05391 [hep-th]} \BibitemShut
  {NoStop}%
\end{thebibliography}

%

\onecolumngrid
\begin{center}
\textbf{\large End Matter}
\end{center}
\twocolumngrid

\textit{The Newman-Unti coordinates}. The NU gauge is based on a family of null hypersurface which can always be introduced in a normal hyperbolic Riemannian manifold \cite{Newman:1961qr,Newman:1962cia}. Suppose this family of null hypersurface is parametrized by $u=$const, and $u$ is chosen as one of the spacetime coordinates. The normal vector to those null hypersurface $\ell=\td u$ must be null. Consequently, $\ell$ is tangent to null geodesics which lie within the null hypersurfaces. 

The remarkable feature of foliating spacetime with null hypersurfaces is that one can formulate the Einstein equation or other perturbative equations on a background spacetime as a characteristic initial value problem \cite{Winicour:2001imp}. More specifically, the equations of motion can be organized into two groups: evolution along $\ell$ on a single null hypersurface, and evolution between different hypersurfaces. It is very convenient to choose the geodesic parameter (e.g., $r$) as one coordinate of the spacetime, so that the evolution equations along a single null hypersurface reduce to ordinary differential equations, because the directional derivative along $\ell$ is proportional to $\frac{\p}{\p r}$. If $r$ is an affine parameter for the geodesic, $\ell=-\frac{\p}{\p r}$. The remaining two coordinates $x^a$ label the geodesics on each null hypersurface. This completes the construction of the celebrated Newman-Unti gauge with the metric assumptions,
\be\label{NU}
g^{uu}=0,\qquad g^{ur}=-1,\qquad g^{ua}=0.
\ee
It is important to note that an affine parameter is not required to formulate the equations of motion as a characteristic initial value problem. In the seminal work of Bondi and collaborators \cite{Bondi:1962px}, a luminosity parameter was employed where the metric component $g^{ur}$ is left free while the determinant of the metric $g^{ab}$ is fixed. This choice would simplify the evolution equations on the null hypersurface, see, e.g., \cite{Chesler:2013lia,Geiller:2022vto,Geiller:2024amx}.


\textit{Minkowski with sphere boundary in generic dimensions}. In four dimensions, the supertranslated Minkowski spacetime with a sphere boundary was obtained in \cite{Compere:2016jwb} by applying a combined supertranslation and Weyl rescaling to the plane boundary Minkowski metric. In this section, we will apply the characteristics to obtain the supertranslation for Minkowski spacetime with a sphere boundary in generic dimensions. The line-element is given by
\be
\td s^2=-\td u^2 - 2 \td u \td r + r^2 \gamma_{ab}(x^c)\td x^a \td x^b,
\ee
where the Latin characters denote indices on the $d$-dimensional sphere. The null condition for the scalar function reads
\be
2\p_r f = (\p_r f)^2 + \frac{1}{r^2} \gamma^{ab}\p_a f \p_b f,
\ee
where $\p_a=\frac{\p}{\p x^a}$. The above equation can be rewritten as
\be
F(r,x^a,p_r,p_a)=p_r^2-2p_r+\frac{1}{r^2} (p_a)^2=0,
\ee
where we define $p_r=\p_r f$ and $p_a=\p_a f$. The characteristic equations for deriving the supertranslation are
\be
\begin{split}
&\frac{\td r}{\td s} =2(p_r-1),\quad \frac{\td{x}^a}{\td s} =\frac{2}{r^2} p^a, \quad \frac{\td {p_r}}{\td s} =\frac{2}{r^3} (p_a)^2,\\
&\frac{\td p_a}{\td s}= - \p_a \gamma^{cd} p_c p_d,\qquad \frac{\td f}{\td s}=\frac{\td r}{\td s} p_r + \frac{\td{x}^a}{\td s} p_a.
\end{split}
\ee
We can deduce the conservation of total angular momentum from the above characteristic equations, $\frac{\td}{\td s} (p_a)^2=0$, which will significantly simplify the derivation of supertranslation. Let $(p_a)^2=L^2$ on the characteristic curves. Then, the constraint equation gives
\be
p_r=1\pm\sqrt{1-\frac{L^2}{r^2}}.
\ee
We will take the minus solution for the outgoing case with initial data given at infinity $r\to\infty$. The solution for the supertranslation is obtained as
\be
f= f_0 ({\bar x}^a) + r \left (1 - \sqrt{1-\frac{L^2}{r^2}}\right), 
\ee
where $L^2={\bar\gamma}^{ab}\p_{{\bar x}^a} f_0 \p_{{\bar x}^b} f_0$. Here, $f_0$ represents the supertranslation at infinity, and ${\bar x}^a$ can be interpreted as new coordinates for the Minkowski spacetime with a supertranslation. The explicit coordinate transformation can be obtained with the precise form of the angular metric $\gamma_{ab}$. 


\blue{\textit{Supertranslation generators from light-ray integral}. We adopt the light-cone coordinates $(x^+,x^-,x_\bot)$, with $x^+$ as the evolution (null) time  that labels null hypersurfaces and $x_\bot=(x_1,x_2)$ denoting the transverse directions. The line-element of Minkowski spacetime is
\be\label{old}
\td s^2=-2 \td x^+ \td x^- + \td x_i \td x_i,
\ee
where transverse indices $i=1,2$ are contracted with $\delta^{ij}$.
}

\blue{A supertranslation is defined from the transition to a new family of null hypersurface labeled by $\bar x^+=x^+ + f(x^-,x_\bot)$, where $f(x^-,x_\bot)$ can be fixed from the characteristic flow as
\be\label{transfer}
\begin{split}
&x^+=\bar x^+ - f(\bar x_\bot) + \frac{1}{2} x^- (\bar\p_i f)^2 ,\\
& x^-=\bar x^-,\qquad x_i=\xbar x_i - x^- \bar\p_i f(\bar x_\bot),
\end{split}
\ee
where we choose the boundary at the finite-distance null hypersurface $x^-=0$ and $\bar\p_i=\frac{\p}{\p \bar x_i}$. In these coordinates, the Minkowski line-element becomes
\be\label{super}
\begin{split}
\td s^2=&-2 \td \bar x^+ \td \bar x^- + \td \bar x_i \td \bar x_i  - 2 \bar x^-  \bar\p_i\bar\p_j f \td \bar x_i \td \bar x_j \\
&+ (\bar x^-)^2 \bar\p_k \bar \p_i f \bar\p_k \bar\p_j f \td \bar x_i \td \bar x_j,
\end{split}
\ee
where transverse indices $i,j,k$ are contracted with $\delta^{ij}$. This result reproduces the gravitational memory induced by impulsive gravitational waves \cite{OLoughlin:2018ebk}. The bulk supertranslation is therefore obtained by propagating the supertranslation defined on a finite-distance null hypersurface along the characteristic flow.
}

\blue{
In light-cone quantization \cite{Brodsky:1997de,Heinzl:2000ht}, the generator of an infinitesimal diffeomorphism $\xi_\mu$ is defined from the energy-momentum tensor as
\be
Q_\xi = \int_\Sigma \td\Sigma_\mu\, T^{\mu\nu} \xi_\nu,
\ee
where $\td\Sigma_\mu$ is the hypersurface measure. In particular, the Hamiltonian and longitudinal momentum are associated to translations along $x^+$ and $x^-$ directions on  $\Sigma$,
\be
P_+=\int_\Sigma \td x^- \td^2 x_\bot T^{+-},\quad
P_-=\int_\Sigma \td x^- \td^2 x_\bot T^{++},
\ee
where $\td\Sigma_\mu=\td x^- \td^2 x_\bot \ell_\mu $ for the hypersurface $\Sigma$.}

\blue{
For supertranslation, the generator is not merely a diffeomorphism associated to change of coordinates. Physically, a supertranslation shifts the null hypersurface, changing the normal vector from $\ell=\td x^+$ to $\bar\ell=\td \bar x^+$. Thus, the supertranslation information must be captured by the generator associated to $\bar\ell$ on $\overline{\Sigma}$,
\be
\overline P=\int_{\overline\Sigma} \sqrt{\bar q_{ij}}\,\td \bar x^- \td^2 \bar x_\bot \overline T^{++},
\ee
where $\bar q_{ij}$ is the induced metric on $\overline{\Sigma}$. Since $\bar\ell$ is null, it lies on the null hypersurface. Hence, $\overline P$ induces the translation along the longitudinal direction and captures the effect from a supertranslation along characteristic curves on the null hypersurface. }

\blue{
If the two hypersurfaces are infinitesimally separated, namely $f\to \epsilon f$ with $\epsilon$ infinitesimal, the coordinate transformation in Eq.~\eqref{transfer} is invertible and a pull-back map from $\bar{\Sigma}$ to $\Sigma$ can be defined. Expanding in $\epsilon$, the supertranslation generator $\overline P$ can be expressed on $\Sigma$ as
\be\label{zero}
{\overline P}=P_- + 2\epsilon \int_\Sigma \td x^- \td^2 x_\bot  \p_i f(x_\bot) T^{+i} + \cO(\epsilon^2).
\ee
The transverse derivative $\partial_i f$ arises from the coordinate transformation, namely the inverse relation of Eq.~\eqref{transfer} at order $\epsilon$, while the change of the hypersurface element contributes only at order $\epsilon^2$.
}

\blue{
We propose that this pull-back operator in Eq.~\eqref{zero} generates an infinitesimal deformation of the null hypersurface associated to supertranslation and acts on the physical states of the theory through the energy–momentum tensor. In this sense, it acts as a light-ray operator that shifts the null foliation and induces a nontrivial transformation on the physical state.
}

\blue{
Equivalently, the pull-back operator can be organized as the charge associated with an infinitesimal diffeomorphism $\xi_\mu$, with $\xi_i=\epsilon \p_i f$. The algebra involving supertranslation generators then follows from the Lie algebra of the vector fields $[Q_{\xi_1},Q_{\xi_2}]=Q_{[\xi_1,\xi_2]}$.
It is straightforward to verify that the supertranslation generator commutes with both $P_+$ and $P_-$, while the commutator between two supertranslation generators appears only at order $\epsilon^2$, so that the supertranslation algebra is Abelian at the leading order. To recover the full algebra between supertranslation generators, one needs to trace more orders in $\epsilon$, which we leave for future investigations \cite{mao}.
}

\end{document}